**Experimental characterization and numerical modelling of the translaminar fracture of woven-ply hybrid fibers reinforced thermoplastic laminates**

C. Bouvet, B. Vieille, J-D. Pujols-Gonzalez

1: Groupe de Physique des Matériaux, Normandie Univ, UNIROUEN, INSA Rouen, CNRS

Avenue de l'Université, 76801 Saint Etienne du Rouvray Cedex, France

2: *Université de Toulouse, Institut Clément Ader, ISAE-SUPAERO – UPS – IMT Mines Albi – INSA – CNRS - 10 av. E. Belin, 31055 Toulouse cedex 4, France*

Address correspondence to: benoit.vieille@insa-rouen.fr

**Abstract**

This study was aimed at investigating the influence of initial notch orientation on the translaminar fracture of woven-ply hybrid fibers reinforced thermoplastic polyether ether ketone (PEEK) laminates. This work is based on the experimental characterization of translaminar fracture of Single-edge-notch bending (SENB) specimens with two initial notches (0° and 45°). Such geometry results in a complex stress state within the laminates plies as well as simultaneous tension/compression failures. A digital image analysis technique has been implemented to monitor the crack initiation and growth during mechanical loading. To better understand the role played by the initial notch orientation as well as the plies orientation contribution to fracture behavior, a specific Finite Element mesoscale modelling was built to account for the deformation mechanisms (namely local plasticity) and the different damage behaviours (fiber breakage in tension and compression, kinking/crushing in compression, delamination) occurring within the plies of quasi-isotropic laminates. Linear elastic fracture mechanics concepts have been applied to quantify the critical translaminar fracture toughness (about 40 kJ/m² in both cases). Finally, the G-R curves were computed from the compliance method to investigate the influence of the initial notch orientation on the evolution of the fracture energy.

## 1. Introduction

In service conditions, a “binary” elasticity-fracture approach to design is not sufficient as damage (usually cracking) may occur in a localised and non-critical manner. It is then necessary to know how this damage is initiated and propagated within materials. The growth of damage will potentially be countered/facilitated by local mechanisms (architecture, microstructure, manufacturing defect, presence of several phases) which strongly depend on the nature and properties of the material and especially on the stress conditions (mechanical, thermal, chemical, environment).

Edge cracks may initiate from notches at the surface of composite laminates due to processing (consolidation, machining) or stress concentrations due to the geometry of composite structures [1]. They may result in the failure of composite structures subjected to service conditions (monotonic or cyclic loadings). It is therefore necessary to evaluate the influence of prominent factors (notch orientation, stacking sequence, matrix ductility, service temperature) on the overstress distribution at the notch tip in order to better understand the subsequent damage mechanisms and failure modes (opening or mixed-mode). In orthotropic or quasi-isotropic composite laminates, transverse matrix cracking and fibers breakage (also known as translaminar failure modes) are usually the primary damage mechanisms occurring in the early phase of mechanical loading. A comprehensive review of techniques for the experimental characterization of the fracture toughness (FT) associated with the translaminar failure modes of continuously reinforced laminated composites is presented in [2]. Depending on matrix ductility, the localized matrix plasticization is an energy dissipative process ruling both fracture behaviour and composite toughness, as these matrix-rich regions may act as cracks barriers and subsequent propagation in laminated fiber-reinforced plastics [3]. It is therefore expected this mechanism to reflect on material toughness measurements either in fiber-dominated laminates (quasi-isotropic) or matrix-dominated laminates (angle-ply). Most references available in the literature deal with tensile or bending specimens with 0° oriented initial notches [4-5] or 45° oriented notch [6-10]. The earliest study was proposed in the mid-seventies by Spencer et al. to investigate the effect of notch and fiber orientation on the crack propagation in FRPs [11]. In woven-ply glass/epoxy laminates, Kaya et al. have shown that the values of FT obtained in Mode I are 78 % and 211 % higher than in mixed mode (Mode I/II) at the same temperature and deformation rate, respectively [6]. More recently, Khaji et al. have used the concept of the maximum strain energy rate criterion to investigate the crack initiation angle for a composite material subjected to a mixed-mode I/II loading. They have concluded that the crack initiation angle and damage factor depend only on the mechanical properties of the material [9]. To investigate the influence of laminates' stacking sequence, Vieille et al. have conducted tensile tests on 5-harness satin weave carbon fabric reinforced PolyPhenylene Sulphide (PPS) laminated composites with an initial edge notch in quasi-isotropic and angle-ply laminates [12]. Depending on the initial notch orientation (0 or 45°), the total strain energy release rate in quasi-isotropic laminates is 6 times as low as in $[+/-45°]_7$ laminates, suggesting that large plastic deformation (due to a matrix-driven behavior and an exacerbated matrix ductility at $T > T_g$) is instrumental in dissipating a great portion of the mechanical energy.

### 1.1 Assessment of fracture toughness in anisotropic materials

Depending on the matrix nature and the laminates' stacking sequence, the fracture mechanics analysis usually depends on the type of the fracture response (brittle or ductile) observed in composite laminates [13]. Different approximations were proposed in the literature for measuring the mode I stress intensity factor $K_I$, which is classically estimated by analytical functions for single-edge-notched specimens loaded in tension or bending [14-15]. In the case of bending specimens with an oblique initial notch, critical fracture toughness values have been proposed by Fett et al. from approximate weight functions to account for mixed-mode (I+II) fracture [16]. To characterize the fracture behavior of materials, the knowledge of the so-called G-R curves is utmost important for design purposes as it represents the evolution of the strain energy release rate along with the macroscopic crack length. The computation of the fracture energy primarily and corresponding data reduction methods depend on the fracture nature (brittle or ductile) as was detailed in [2]. These curves are usually computed from the compliance loss $C$ (corresponding to the displacement over load ratio at any time of the mechanical testing) and the corresponding gradual crack growth in agreement with the ASTM standard test method E1820 [17-21]. In accordance with this ASTM standard, the use of LEFM requires a linear load-displacement behavior and very localized plastic deformation at the crack tip [17]. The relationship between the compliance and the crack length $a$ is classically associated with the calculation of the strain energy release rate $G$ by the following expression:

$$G = \frac{P^2}{2b}\frac{dC}{da} \quad (1)$$

In Eq. (1), $b$ is the thickness of the laminate and $P$ is the applied force. From Eq. (1), it is therefore possible to compute the strain energy release rate in mode $I$ and mixed-mode. The crack growth is usually evaluated by means of a digital images' analysis described in section 2.2.2.

### 1.2 Numerical mesoscale damage models

Compared to the number of reported numerical models for unidirectional composites [22-30] very few woven-ply damage models have been developed [21, 31-33]. Several numerical models to predict damage in unidirectional fiber reinforced thermosetting (TS) composites, more specifically in unidirectional carbon-fiber reinforced epoxy resin composites, are available in the literature [25-26, 28-30]. Experimental studies have shown that woven fabric reinforced composite structures exhibit higher interlaminar strengths and damage tolerance than unidirectional composite structures, what motivated the growing interest in woven composites. As with unidirectional composites, most numerical woven damage models have been developed for woven-ply carbon fiber-reinforced epoxy resin composites [21, 31], but models have been recently developed to simulate the behavior of woven-ply carbon fiber-reinforced composites consisting of high-performance thermoplastic (TP), PEEK and PPS resin [32-33]. The increasing utilization of woven composites requires the development of corresponding numerical models, due to the complex and multi-scale nature of the damage formation in laminated composites. In the literature, there is no unique model applicable to a variety of loading conditions and geometries [34]. There are very few experimental and numerical studies dealing with the fracture behavior of woven fabric reinforced thermoplastic composites. Liu et al. [32] have developed a mesoscale damage model for predicting damage in woven carbon-fiber-reinforced PEEK prepreg composites under compressive loading. Jebri et al. [33] have tested three-dimensional finite element model implemented in Abaqus/Explicit to reproduce the mechanical behavior of 5-harness satin weave carbon/PPS composite with different fiber orientations (warp, diagonal and weft) under tension. Fiber failure and matrix cracking were modeled using the Hashin failure criterion and a damage evolution law. Delamination was simulated using cohesive zone elements between composite layers. The non-linear shear response was considered using an exponential function, where the constants were obtained by a tensile test of the laminate oriented at 45°. The damage evolution was calculated from classical strain-stress relationships and the fracture toughness associated with the fiber failure in tension and compression. A good agreement was achieved between the experimental and simulation results in terms of load–displacement curves and the presence of a centered hole affects the translaminar fracture as fiber breakage grows from the vicinity of the hole to the laminate edges. Using a combination of interlaminar and intralaminar damage models implemented in Abaqus/Explicit [35], it is possible to capture different composite failures modes, such as fiber fracture and matrix cracking as well as delamination. A strain-based damage criterion was employed by the authors to evaluate the fiber failure in warp/weft directions, due to tensile and compressive loading respectively. This criterion combines the ultimate strains of the fiber to the translaminar fracture toughness, strength and characteristic length of the element. The translaminar fracture toughness values associated with fiber-dominated tensile and compressive failure are 109 $kJ/m^2$ and 52 $kJ/m^2$ respectively.

**1.3 Objectives of the study**

This work is aimed at investigating the fracture behavior of woven-ply reinforced laminated composites consisting of highly ductile thermoplastic matrix (PEEK) and brittle fibers (carbon and glass woven fabrics) by means of a finite element model implemented into the Finite Element (FE) code Abaqus/Explicit [35]. One of the main objectives of this study is to test the capability of a numerical model to predict the translaminar failure based on SENB tests. These tests are more difficult to analyze due to the combination of several physical phenomena at the same time (failure in tension and compression), which involve kinking, crushing, resin plasticity and fiber breakage. In particular the adopted numerical approach assumes that only a fiber failure in mode I is possible, and then only a mode I fracture toughness of fiber failure is necessary to simulate crack propagation in mixed mode I/II. In fact, the plasticity of the resin under in-plane shear loading is so important that this plasticity smooths the shear stress and avoids the mode II fiber failure. The effect of local stress state on the fracture behaviour is specifically addressed by means of SENB specimens with two different initial notch orientations (0° and 45° oriented notches). Finally, the G-R curves obtained from experimental characterization (compliance method) and numerical modelling were compared.

## 2 Materials and methodology

### 2.1 Materials and specimens

The consolidated laminates consist of 14 inner woven carbon-PEEK plies and two outer woven glass-PEEK plies provided by Toho Tenax [13, 36]. The two-outer glass-PEEK plies are considered for electrical protection purposes. The carbon and glass fiber woven fabrics are balanced in the warp and weft directions. The PolyEther Ether Ketone (PEEK) matrix has a glass transition temperature $T_g$ = 143°C. The laminated plates obtained by thermo-compression are made up of carbon–PEEK 5HS woven plies prepregs of 0.31 mm-thickness and glass–PEEK prepregs of 0.08 mm-thickness. The stacking sequence of laminates is nearly quasi-isotropic: $[(0/90)_G,[(0/90),(\pm45)]_3,(0/90)]s$ (with G index for glass fibers ply). The mechanical properties of the quasi-isotropic laminates are specified in Table 1. One may specify that this draping sequence is not virtually quasi-isotropic because there are more plies in the 0° direction than in the 45° direction. The bending test specimens were cut by water jet from 600 × 600 $mm^2$ plates. SENB specimens are characterized by two different orientations of the initial notch (Fig. 1) in order to observe a purely mode I failure (0° oriented notch) or a mixed mode failure (45° oriented notch). The 0° SENB specimens are characterized by a ratio $a/w = 0.33$, whereas the 45° SENB specimens have a ratio $a/w = 0.48$ so that the projected area is the same. Notches were machined by means of a precision endless diamond wire saw whose radius is 0.085 mm. Three specimens were tested in each configuration. The average specimen's thickness (4.5 mm) is significant enough to consider the sample as thick structures in which a plane strain state prevails.

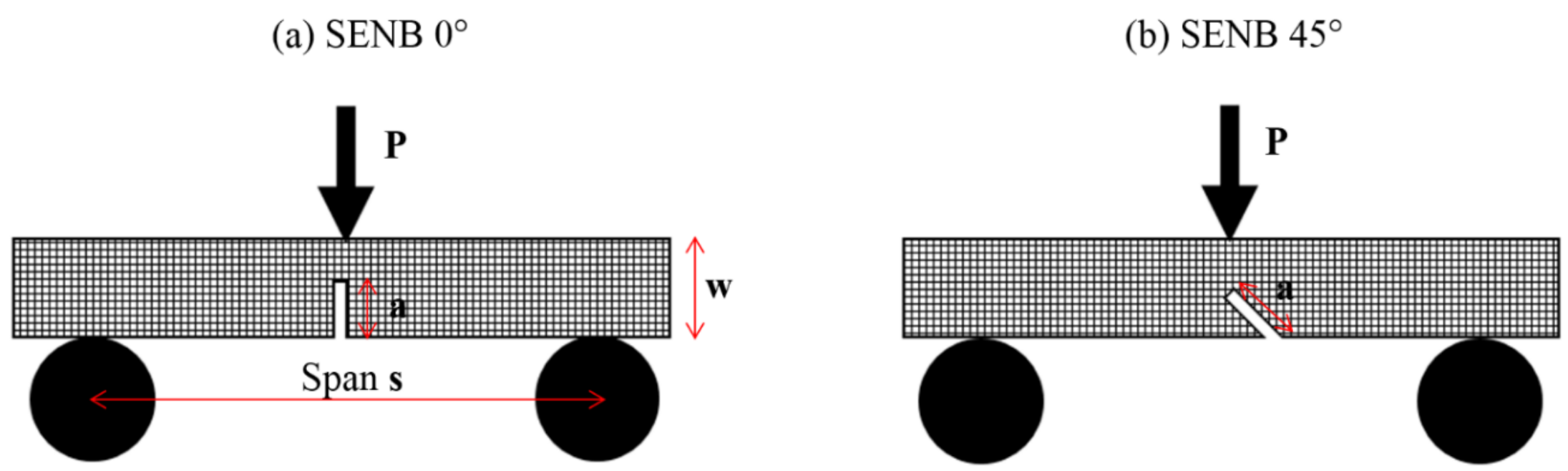

Figure 1 – Single-Edge Notched bending specimens: (a) 0° initial notch – (b) 45° initial notch

## 2.2 Methodology

### 2.2.1 Mechanical testing

SENB specimens were subjected to monotonic bending loadings at room temperature, by means of a MTS 810 servo-hydraulic testing machine equipped with a 100 kN capacity load cell and a thermal chamber. The mechanical properties in bending were determined according to the European standards EN 2562 [37]. The bending strain and stress are defined as follows:

$$\varepsilon_{bending} = \frac{6dw}{s^2} \quad \text{and} \quad \sigma_{bending} = \frac{3Ps}{2bw^2} \tag{2}$$

Where $b$ is specimen thickness, $w$ is the specimen width, $s$ is the span between the support points (80 mm), and $d$ is the displacement applied to the upper part of the specimen (Fig. 1).

### 2.2.2 Full-field measurements

A two-dimensional Digital Image Correlation (DIC) technique is used to measure full-field strains. A high-speed monochromic camera Phantom Miro M310 records digital images during loading. The acquisition frequency is 3,200 frames-per-second at full resolution (1280 × 800 pixels). The Green-Lagrange strain field is computed from the 2D displacement field by means of the VIC-2D correlation software (provided by the company Correlated Solutions). The use of DIC requires a high intensity source combined with a fiber optic light guide (Fig. 2). The DIC technique compares the digital images for various small regions (known as subsets) throughout the images before and after deformation using fundamental continuum mechanics concepts, locating the positions of each of these subsets after deformation through digital image analysis.

(a) (b)

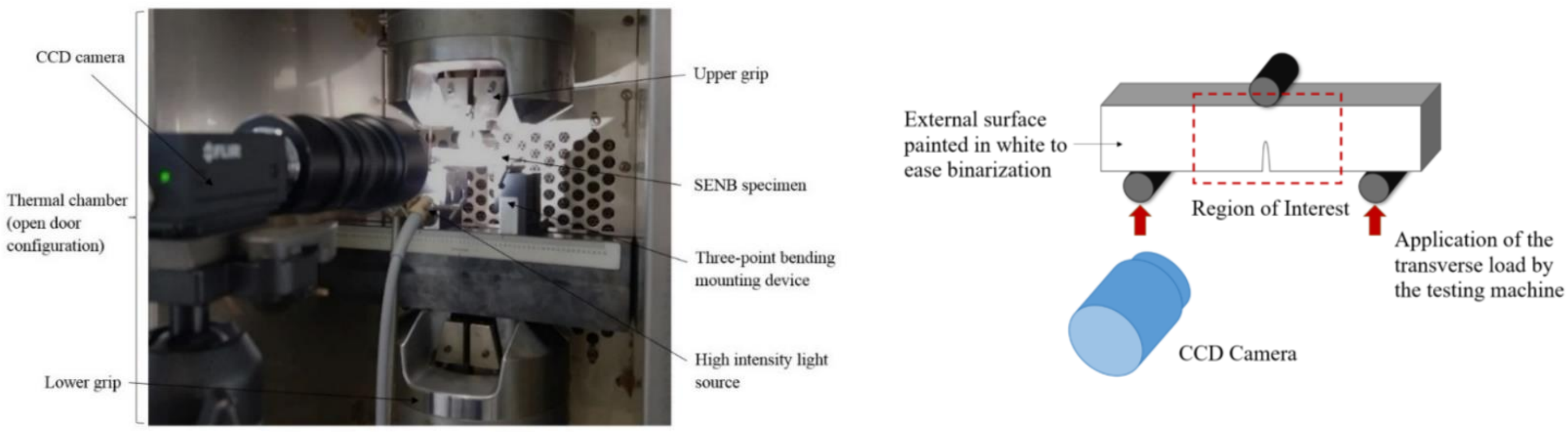

Figure 2 – Monitoring of the crack growth by means of a digital images analysis technique: (a) experimental set-up – (b) schematic representation of the SENB specimen

When a material discontinuity appears in the pattern, the homogeneous linear mapping in this region does not present deformed surface and the small subset position is not computed. Using an algorithm implemented in the Scilab code, the crack tip is located via the discontinuity of the strain field resulting from the crack propagation. Catalanotti et al. used a similar technique based on the displacement discontinuity within the pattern [38]. Scilab is a free and open-source, cross-platform numerical computational package and a high-level, numerically oriented programming language [39]. The crack length is then determined from the crack tip position in each pattern, regardless the orientation of crack within the pattern. The spatial resolution of the crack length is defined by the size of the subsets used in the DIC method.

### 2.2.3 Numerical modelling of the translaminar cracking: Principle and description

A numerical model consisting of a failure criterion based on fracture mechanics is tested to account for the dissipation of the critical strain energy release rate in opening mode (mode I) due to fiber breakage. This model is implemented in user-defined VUMAT subroutines and runs with Abaqus/explicit solver [35,40]. A focus on translaminar cracking and damage evolution in woven-ply laminated composites is proposed, along with the simulation of SENB tests. The damage model is based on ply failure mechanisms at a ply level.

## Meshing strategy

A homemade mesh maker was developed in order to create a mesh complying with fibers directions. This point is a key point of the proposed model and enables to easily obtain a value of dissipated energy equal to the translaminar FT, taking into consideration the constant finite element length [41]; this specific point will be developed in the next section. Positions of nodes are uniformly stacked in row and column to respect the plies orientations. Square shaped elements C3D8 are used in 0°/90° plies and in ±45° plies in order to comply with the fibers direction and to get coincident nodes between adjacent plies (Fig. 3). Elements wedge shape C3D6 are used to adapt the mesh between adjacent plies. The total number of elements used in these simulations is about 90,000 elements and the maximum computation time is about 24h with 36 CPUs.

The delamination between plies is driven using a contact algorithm and cohesive contact law based on fracture mechanics [35]. The modeling of delamination will be detailed in section III. The size of the square shaped elements used to account for propagation damage is approximately 0.625 mm for the 0°/90° plies and 0.884 mm for the ±45° plies (of course the mesh size of the ±45° plies equals $\sqrt{2}$ times the 0/90° plies one). The average ply thickness of the laminates is 0.31 mm. SENB tests are modeled using the same specifications as the ones used in experimental testing [3]. The bending load is applied by means of semi-spherical analytical rigid cylinders (Fig. 4a).

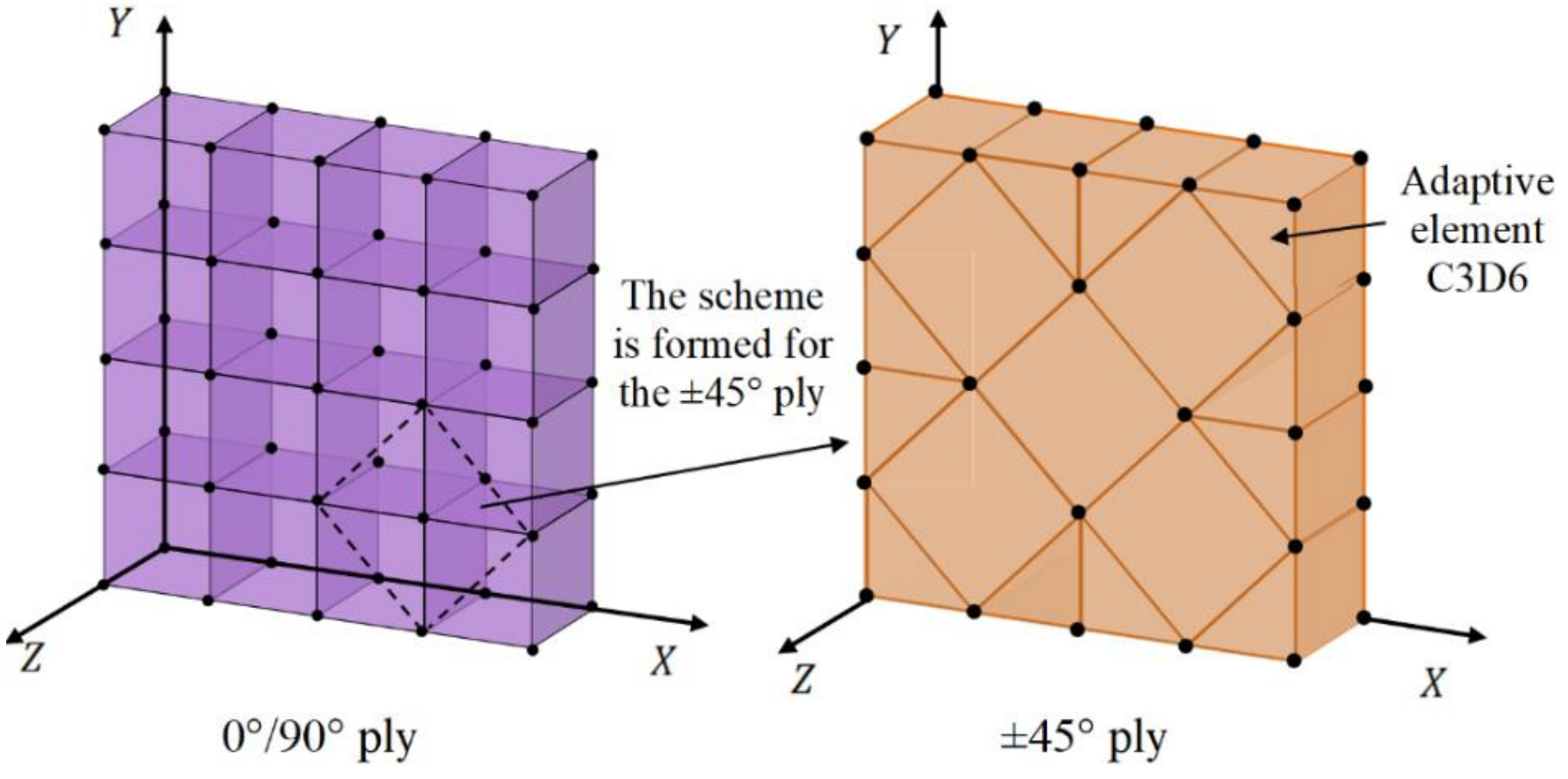

Figure 3 – Mesh principle with oriented elements in the fiber directions

The mesh used in the damaged area being generated from finite elements aligned in the fibers' direction (Fig. 3), it is difficult to mesh the exact shape of the crack. By simply removing a row of elements, the mesh is created in this area (Fig. 4b). Consequently, such meshing of the crack does not make possible to model the crack tip radius. However, it is assumed a second order effect, contrary to the respect of the fibers direction, which is a key issue in the present case. In the Fig. 4b, it can be seen the scheme of meshes of 0° and 45° notch and of 0° and 45° propagation. To account for the crack propagation at 0° or 90° of the fibers direction, a mesh row (propagation at 0° or 90° in 0/90° plies or at ±45° in ±45° plies) is simply removed from the meshing near the crack tip. In the case of a crack propagation at ±45° of the fiber directions, it is more complex as it implies to remove two mesh rows (cracking at ±45° in ±45° plies with the 0° propagation). This point will be further detailed in the next section dealing with the fracture criterion.

(a)

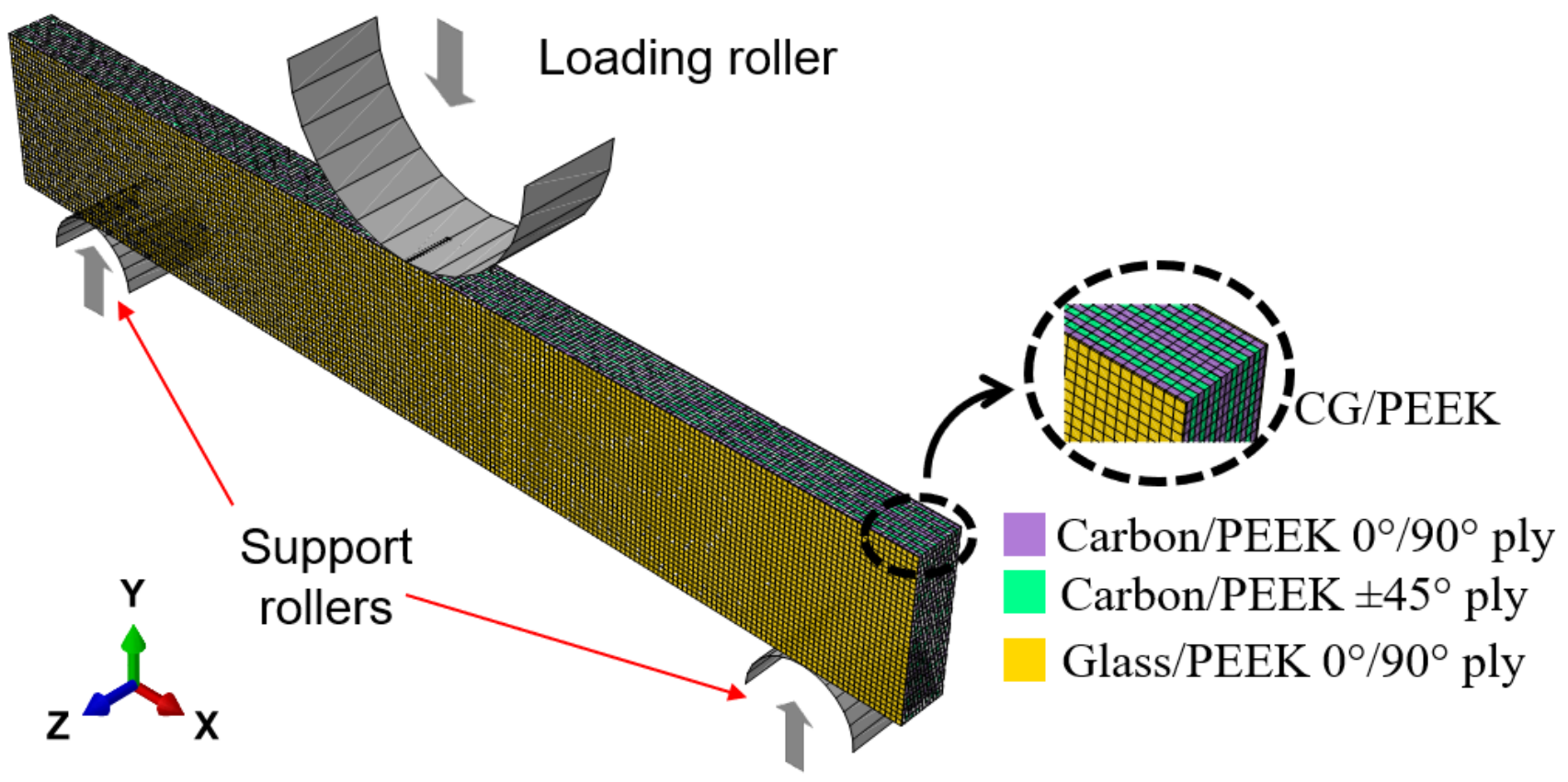

(b)

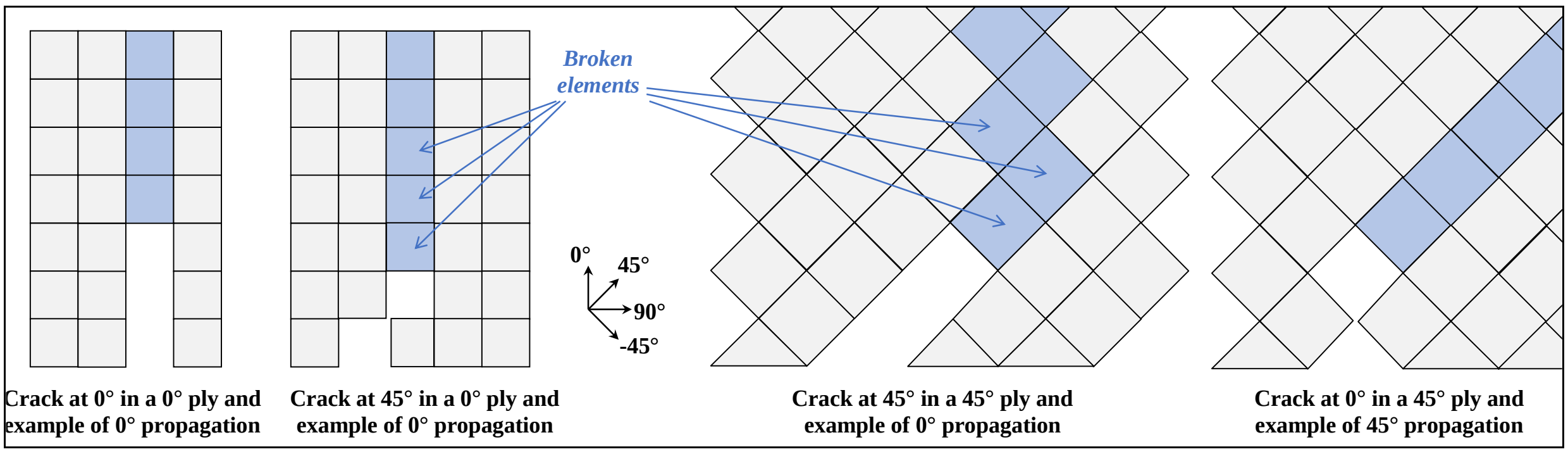

Figure 4 – Mesh and boundary conditions used in the FE modelling of the hybrid CG/PEEK laminates (a) and principle of the crack initial mesh and propagation (b)

Although the studied draping sequence has 16 plies, including the two thin (0.08 mm thick) glass plies, only the 14 carbon plies were considered in the proposed FE model. Indeed, the yarn size is a key factor to be fulfilled in the proposed modelling. Actually, the yarn is 0.67 mm wide, and this width is considered in the ±45° plies. In order to comply with the nodes coincidence between two consecutive plies, the yarn width is √2 times higher in the 0° plies. This mesh size leads to a high size ratio (about 8.5) in the glass plies due to their low thickness value. Such ratio results in hourglass deformation modes (Fig. 5), particularly after the sudden crack propagation. Therefore, the glass plies were not considered in the model. In addition, the FT of glass fibers cannot be determined accurately.

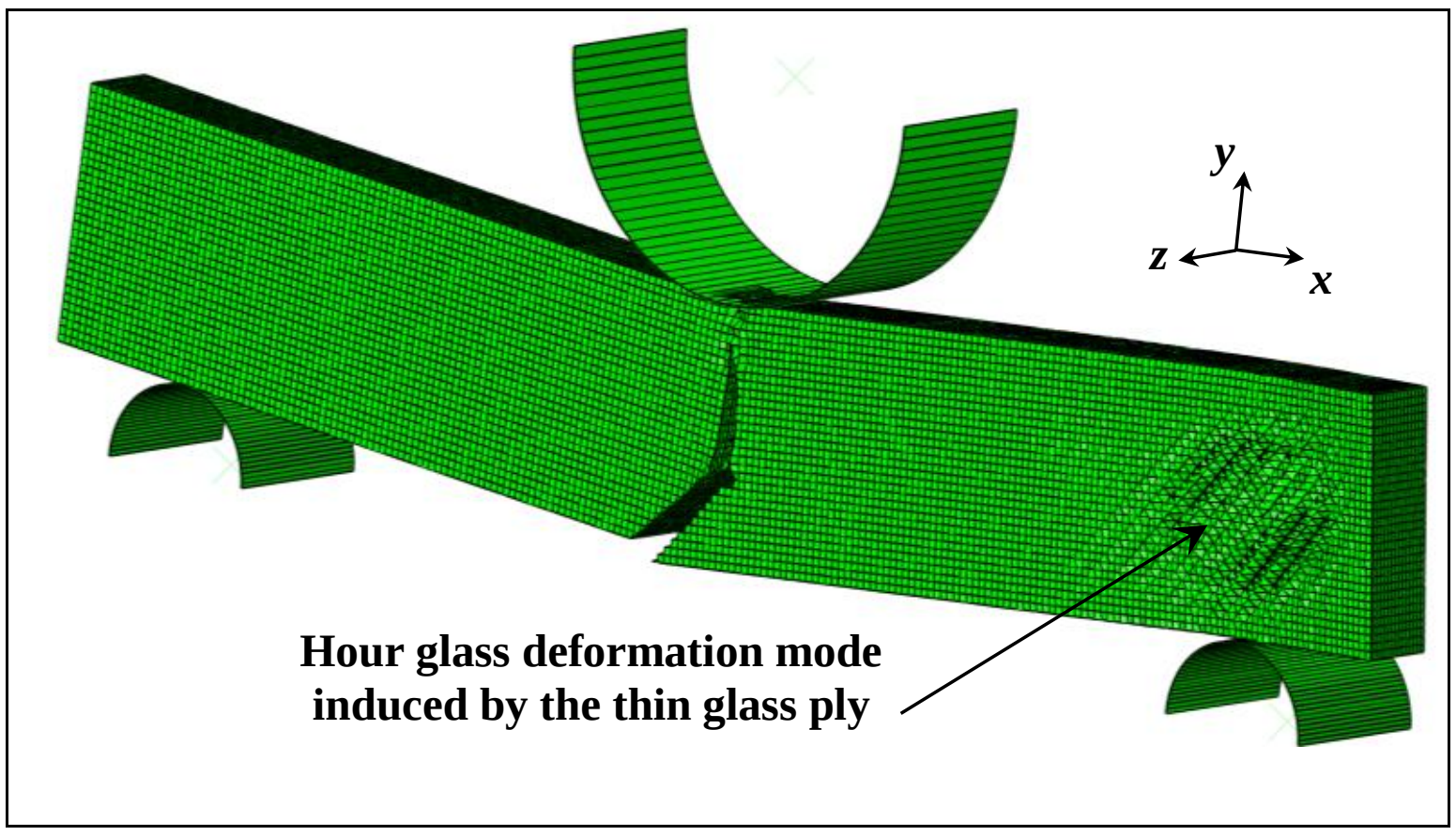

Figure 5 – Hourglass deformation mode resulting from the high size ratio of the glass plies (deformation scale factor = 1 in *x*- and *y*-directions and 100 in *z*-direction)

**Fiber breakage modelling**

SENB tests show that fiber failure and crack growth are primarily associated with the breakage of 0° fiber bundles in tension but also reveal that several physical phenomena (fiber breakage, kinking, crushing) occur at the same time [13] making the translaminar cracking analysis more complex to be simulated [2,22,40,42]. Based on these experimental observations, three main physical phenomena were found to be responsible for dissipating the mechanical energy: fiber failure, in-plane shear plasticity and crushing plasticity. The behavior laws implemented in the present numerical model are based on these physical phenomena.

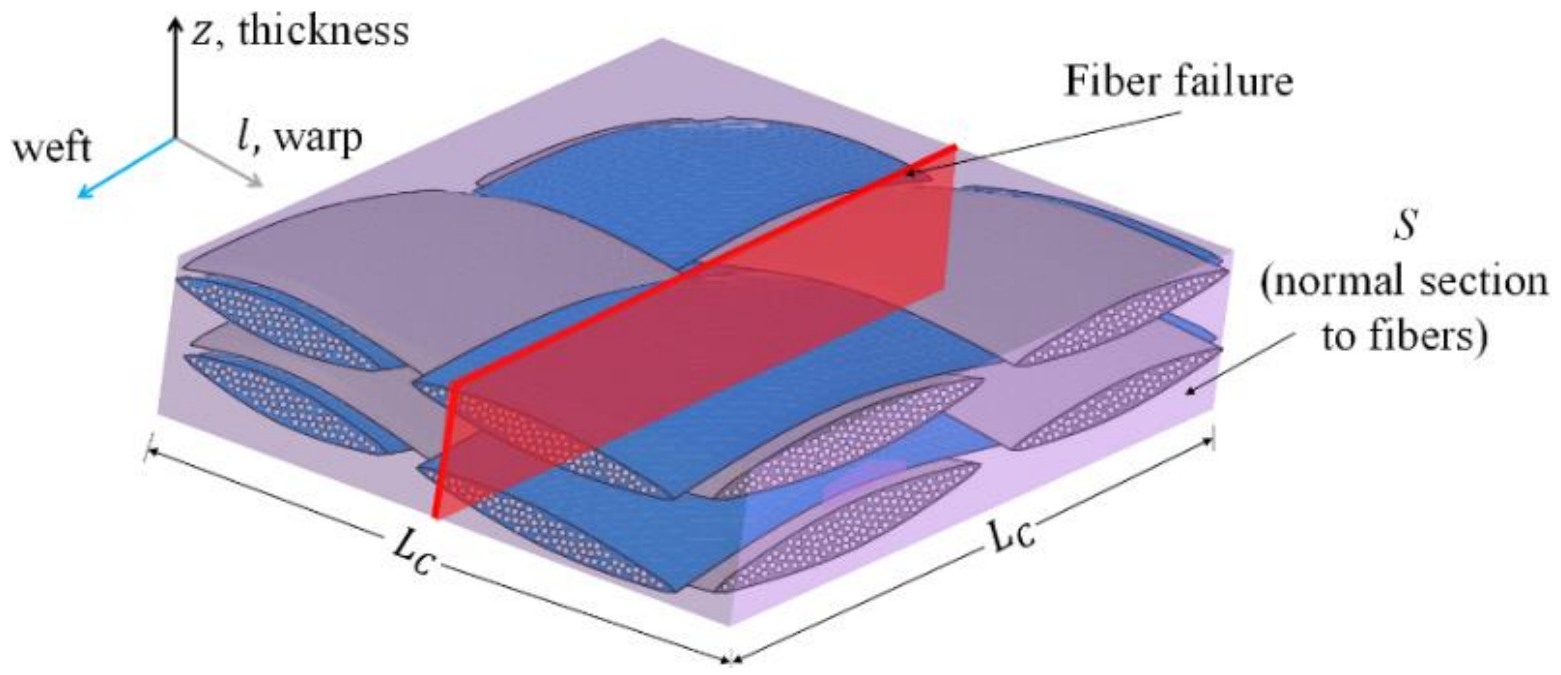

Figure 6 – Representation of fiber breakage at the level of the volume element

Fibers breakage is considered using a failure criterion formulated within volume elements (Fig. 6). This fracture mechanics-based criterion is built to dissipate the critical strain energy release rate in opening mode (*I*) due to fiber fracture in the volume elements regardless the element size [41]. Then a classical formulation between the integration points of the elements provides a constant critical energy release rate per unit area [30]:

$$\int_V \left( \int_0^{\varepsilon_1} \sigma d\varepsilon \right). dV = S.G_f \quad (3)$$

where $G_f$ is the critical strain energy release rate in mode $I$, $\sigma$ and $\varepsilon$ are the stress and strain in fiber direction, and $\varepsilon_1$ is the strain in fiber direction at ultimate failure. As shown in Fig. 7, this formulation is applicable either under tension ($G_f^T$ and $\varepsilon_1^T$) or compression ($G_f^C$ and $\varepsilon_1^C$), both in longitudinal and transverse fiber directions. In compression, crushing phenomenon is also simulated, as further described. $V$ and $S$ are the volume and the normal section of the element (Fig. 6), respectively. The dimensions can be reduced in terms of characteristic internal length $L_C$ to avoid mesh-dependent solutions [41].

In tension, failure is classically driven by a damage parameter, allowing the dissipation of the fracture energy of tensile corresponding to fiber failure ($G_f^T$), regardless the length of the element ($L_C$). In compression, the failure is driven by a plastic strain ($\varepsilon^P$), also allowing the dissipation of the fracture energy associated with compressive fiber failure ($G_f^C$), regardless the length of the element ($L_C$). In addition, once the fracture energy is dissipated in compression, a constant crushing stress $\sigma_{crush}$, is imposed in order to simulate the crushing phenomenon (Fig. 7).

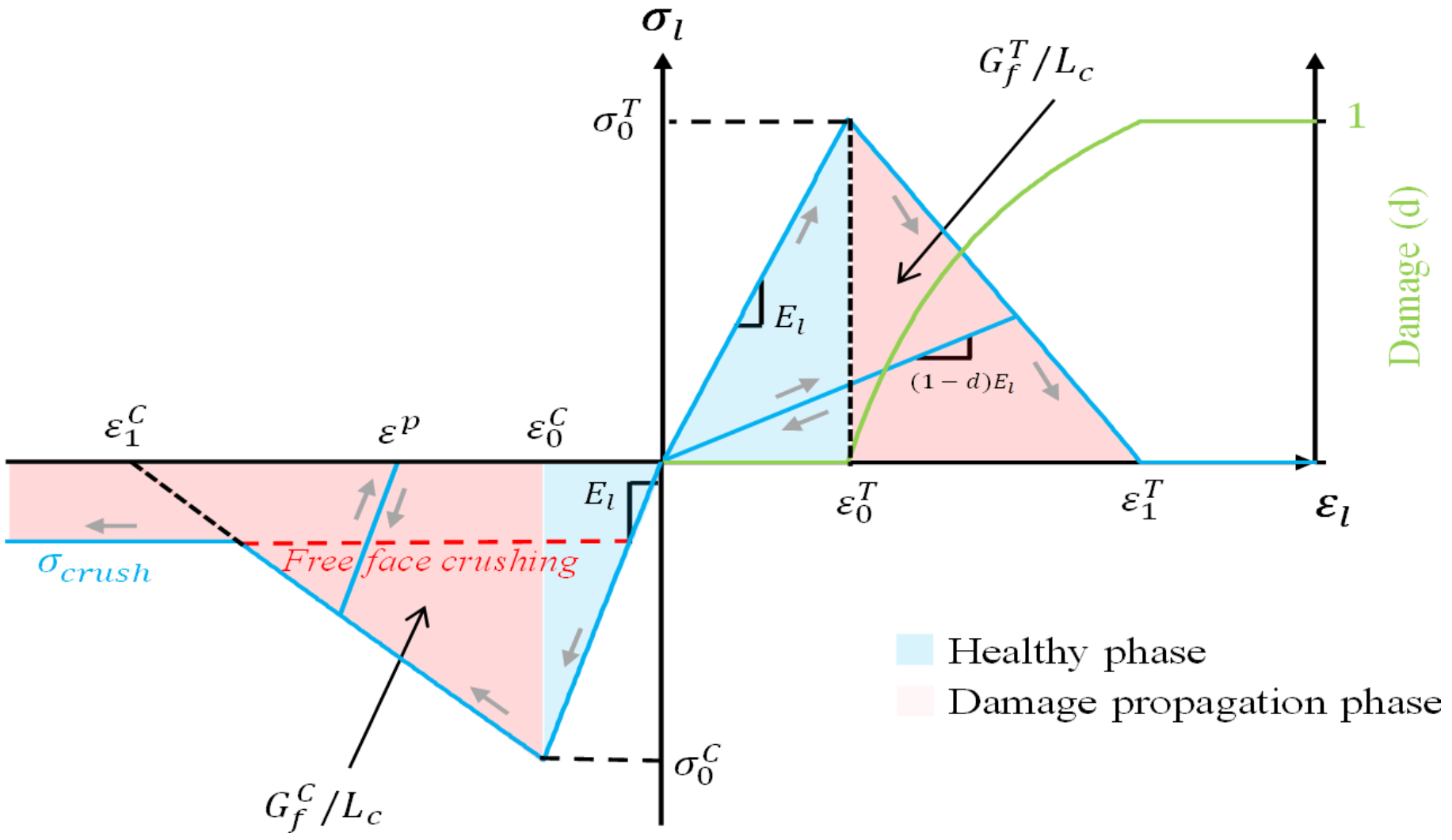

Figure 7 – Overall fiber failure behavior law with damage initiation and damage propagation under tension and compression, in both longitudinal and transverse fiber directions

With respect to longitudinal or transverse directions, when the failure strain in tension ($\varepsilon_0^T$) or in compression ($\varepsilon_0^C$) is reached, a damage variable ($d$) in tension and a plastic strain ($\varepsilon^P$) in compression are used to compute the stress corresponding to the fracture energy dissipation in opening mode $I$. One may notice that the area of the triangle (healthy phase + damage propagation phase) on the stress / strain curve (Fig. 7) represents the ratio of the translaminar FT divided by the characteristic length of the element $G_f^{T,C}/L_c$. Consequently, the strain at ultimate failure ($\varepsilon_1^T$ or $\varepsilon_1^C$) depends on finite element size [41]. Then the tensile stress is calculated using the orthotropic stiffness coefficients as follows:

- In the longitudinal direction, $\varepsilon_l > 0$:

$$\sigma_l = (1-d)[H_{ll}\varepsilon_l + H_{lt}\varepsilon_t + H_{lz}\varepsilon_z] \tag{4}$$

- In the transverse direction, $\varepsilon_t > 0$:

$$\sigma_t = (1-d)[H_{tt}\varepsilon_t + H_{lt}\varepsilon_l + H_{tz}\varepsilon_z] \tag{5}$$

Where $H_{ij}$ is the $ij$ component of the elastic properties' matrix and $d$ is evaluated in order to obtain a linear decreasing of the stress versus strain in dissipating the fracture energy (Eq. 3).

In addition, damage $d$ is considered as strictly increasing. Once the critical energy release in compressive fiber failure is dissipated, a plastic behavior ($\varepsilon^p$) should be considered and a crushing stress $\sigma_{\text{crush}}$ should be applied according to a plateau, in order to simulate the crushing behavior. In order to avoid a discontinuity between the fracture toughness dissipation phase and the crushing phase (Fig. 7), this plastic behavior also contributes to the dissipation of the fracture energy. Thus, the compressive stress is calculated using the plastic strain and the orthotropic stiffness coefficients as:

- In the longitudinal direction, $\varepsilon_l < 0$:

$$\sigma_l = H_{ll}\left(\varepsilon_l - \varepsilon_l^p\right) + (1-d)[H_{lt}\varepsilon_t + H_{lz}\varepsilon_z] \tag{6}$$

and:

$$f_l = |\sigma_l| - \sigma_{\text{crush}} \leq 0 \tag{7}$$

Where $f_l$ is a plastic function and $\sigma_{\text{crush}}$ is the crushing stress. A coupling between plasticity and damage is considered in order to correlate both damage and plastic strains:

$$H_{ll}\left(\varepsilon_l - \varepsilon_l^p\right) = (1-d)H_{ll}\varepsilon_l \quad \Rightarrow \quad d = \frac{\varepsilon_l^p}{\varepsilon_l} \tag{8}$$

This coupling provides a continuity of behavior in compression between the damage zone (the zone where the FT is dissipated) and the crushing zone (the zone of constant stress characterizing the crushing of the material). This coupling implies that the plastic strain $\varepsilon_l^p$ can only decrease (or increase in absolute value), meaning that crushing-induced damage may only increase. A similar law is used in the transverse direction ($\varepsilon_t < 0$):

$$\sigma_t = H_{tt}\left(\varepsilon_t - \varepsilon_t^p\right) + (1-d)[H_{lt}\varepsilon_l + H_{tz}\varepsilon_z] \tag{9}$$

and:

$$f_t = |\sigma_t| - \sigma_{\text{crush}} \leq 0 \tag{10}$$

with a similar coupling plasticity/damage, as in the longitudinal direction:

$$H_{tt}\left(\varepsilon_t - \varepsilon_t^p\right) = H_{tt}(1-d)\varepsilon_t \;\Rightarrow\; d = \frac{\varepsilon_t^p}{\varepsilon_t} \tag{11}$$

Then, a failure criterion is considered to ensure the coupling of the energy dissipation associated with mode *I* in tension/compression and in longitudinal/transverse directions:

$$\frac{G_I^{l,T} + G_I^{t,T}}{G_f^T} + \frac{G_I^{l,C} + G_I^{t,C}}{G_f^C} = 1 \tag{12}$$

This relation means that a failure in the longitudinal direction results in a failure in the transverse direction (and vice versa). This is a strong assumption and should be confirmed using bi-axial failure experimental tests. However, such tests are difficult to perform and no standard test method is currently available. In addition, it is worth noting that it is necessary to break two elements to model the crack propagating at 45° of the mesh (Fig. 4b). This failure criterion stipulates that the sum of the energy dissipated in the longitudinal and in the transverse directions, makes it possible to break one element associated with the breakage of the warp yarn and one element associated with the breakage of the weft yarn; providing a reasonable physical meaning to the numerical model. As was mentioned previously, the specific propagation of cracks at 45° of the fibers direction implies $\sqrt{2}$ times more energy than the FT value required to grow the crack (Fig. 4b). At the same time, the propagation of a crack at 0° of the fibers directions requires only the breakage of one row of elements (Fig. 4b), which is consistent with the physically observed phenomenon, and implies exactly the dissipation of the FT value. To account for the energy dissipated by plastic deformations, the in-plane shear strain ($\gamma_{lt}$) is computed from a plasticity law:

$$\tau_{lt} = G_{lt}(\gamma_{lt} - \gamma_{lt}^{p}) \tag{13}$$

and:

$$f_{lt} = |\tau_{lt}| - \tau_{lt}^{0} - \lambda\left(\gamma_{lt}^{p}\right)^{\eta} \leq 0 \tag{14}$$

Where $\tau_{lt}$ is the in-plane shear stress, $\tau_{lt}^{0}$ is the shear yield stress, $\lambda$ is the strength index, $G_{lt}$ is the in-plane shear stiffness, $f_{lt}$ is the in-plane shear plastic function, $\gamma_{lt}^{p}$ is the plastic in-plane shear strain and $\eta$ is the strain hardening exponent.

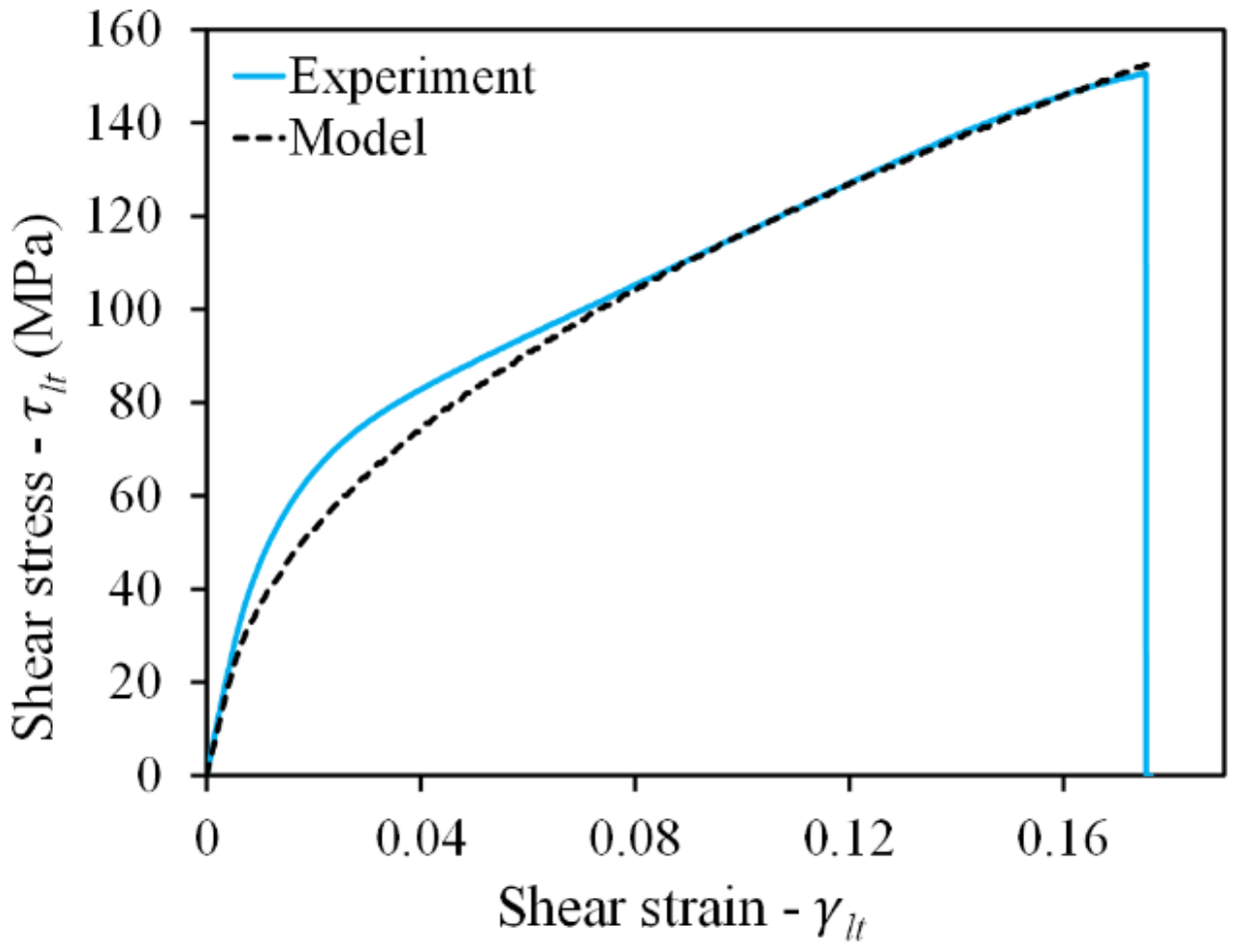

Figure 8 – Experimental and numerical shear stress-strain curve of angle-ply C/PEEK laminate characterized by an elastic-plastic behavior

The elastic-plastic behavior of C/PEEK laminates in shearing is classically identified (Tab. 3) from tensile tests performed on [±45°] specimens (Fig. 8). The normal stress in the out-of-plane direction is classically calculated as:

$$\sigma_z = H_{zz}\varepsilon_z + (1-d)[H_{lz}\varepsilon_l + H_{tz}\varepsilon_t] \quad (15)$$

$$\begin{cases} \tau_{lz} = G_{lz}\gamma_{lz} \\ \tau_{tz} = G_{tz}\gamma_{tz} \end{cases} \quad (16)$$

where $H_{zz}$, $G_{lz}$ and $G_{tz}$ are considered to be damage-independent to avoid the distortion of elements. In addition, this damage is a second order effect compared to the in-plane damage in the $(l, t)$ plane.

**Delamination**

The delamination is simulated by considering the interaction between surfaces with a contact algorithm and cohesive contact law based on a hard pressure-overclosure relationship, which is implemented in the material model's library of Abaqus [35]. A damage criterion allowing the simulation of the degradation and possible breakage of the liaison between adjacent surfaces is also available in this FE code. At first, a quadratic tension-interaction criterion based on the ultimate strengths of the matrix in transverse ($\sigma_t^d$) and shear ($\tau_{lt}^d$) directions are used. Then, damage evolution is defined by a power law fracture criterion based on the fracture toughness of the matrix in mode I ($G_{I_c}^d$) and mode II ($G_{II_c}^d$).

**Numerical model parameters**

The properties used in the simulation are categorized in three parts: elastic properties, plastic properties and fracture properties. All those properties were investigated and identified from experimental tests conducted on the laminated composites studied described in Tab. 2-4 [40]. An initial stiffness value of 500 GPa/mm was chosen for the delamination interfaces [30].

## 3. Results and discussion

### 3.1 Influence of notch orientation on fracture behavior

From the macroscopic response of SENB specimens, it appears that the notch orientation seems to have moderate influence on the mechanical behavior in bending. The stress-strain curves are virtually elastic up to the ultimate strength where the load borne by the specimen significantly drops. Therefore, a quasi-brittle fracture behavior is observed along with a gradual decrease in the load as the macroscopic crack propagates (Figure 9). The failure of both specimens' configurations (with 0° and 45° initial notches) is therefore characterized by a gradual compliance loss. This allows the application of the compliance method to compute (from Eq. 1) the fracture toughness of the CG/PEEK laminates subjected to bending conditions. From the macroscopic mechanical properties' standpoint, it also appears that the ultimate bending strength decreases slightly (-11%) in 45° SENB specimens with respect to SENB specimens with a 0° notch (Table 5). One may recall that the initial notch orientation (0 or 45°) machined in quasi-isotropic SENB specimens is expected to induce primarily a mode I failure in 0° specimens and a mixed-mode (I+II) failure in 45° specimens. This might explain the lower value of ultimate strength in 45° SENB specimens (-11%).

(a) (b)

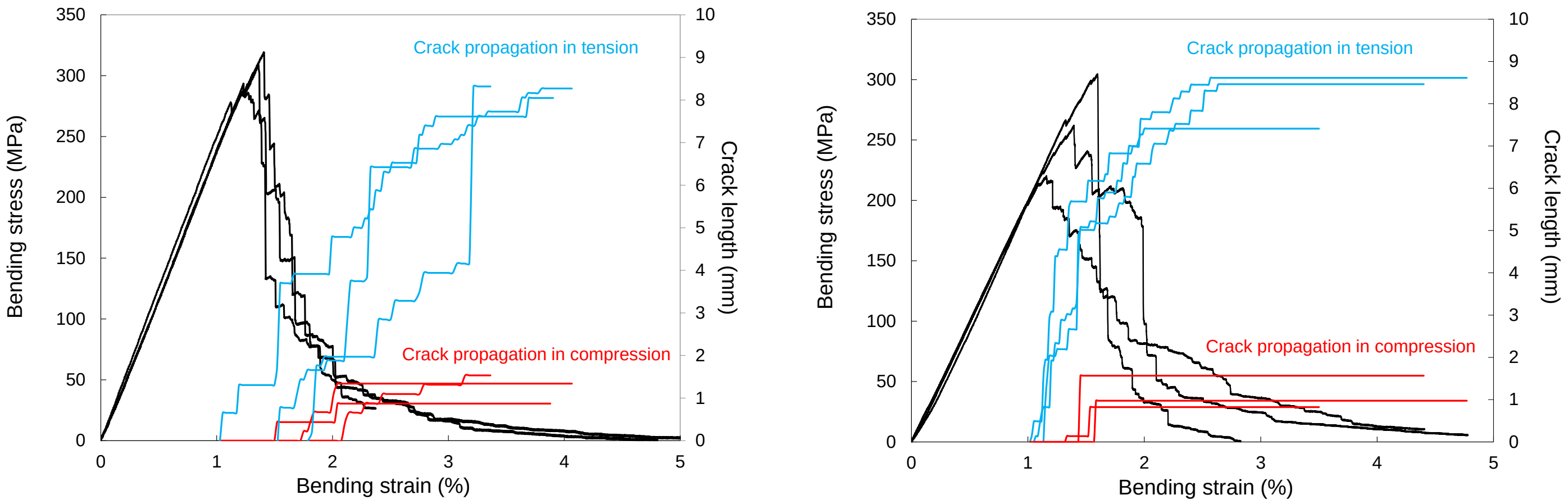

Figure 9 – Influence of initial notch orientation on the experimental response and the evolution of the crack propagation in tension/compression within CG/PEEK laminates subjected to a mechanical loading in bending: (a) 0° notch – (b) 45° notch

The crack initiation is approximately determined from the in-situ observation of the specimen's surface by means of the CCD camera. The digital images analysis introduced in section 2.2.2 allows the monitoring of the crack growth in tension and compression. The strain at initiation is virtually the same in both testing configurations. As the tests were conducted in displacement-controlled mode, it therefore means that crack initiates at about the same time in 0 and 45° specimens (Table 5). The influence of notch orientation on the stress value at initiation is noticeable (-16%) as well. This indicates that the mechanical energy required to initiate the crack is higher in 0° specimens experiencing primarily mode I failure. 45° specimens are characterized by a mixed-mode failure which seems to influence the crack propagation in the early stage of damage. It is also worth noting that crack initiation occurs in the elastic domain of the mechanical behavior.

From the macroscopic bending responses shown in Figure 9, it appears that the gradual decrease in the transverse load $P$ applied to SENB specimens comes along with a stable transverse crack propagation. The transverse crack actually corresponds to simultaneous failure mechanisms: (1) a compressive failure in the upper part of the specimen (the one below the contact area between the upper cylinder allowing the application of the transverse force) and (2) a breakage of 0° fibers in tension near the crack tip (Fig. 10).

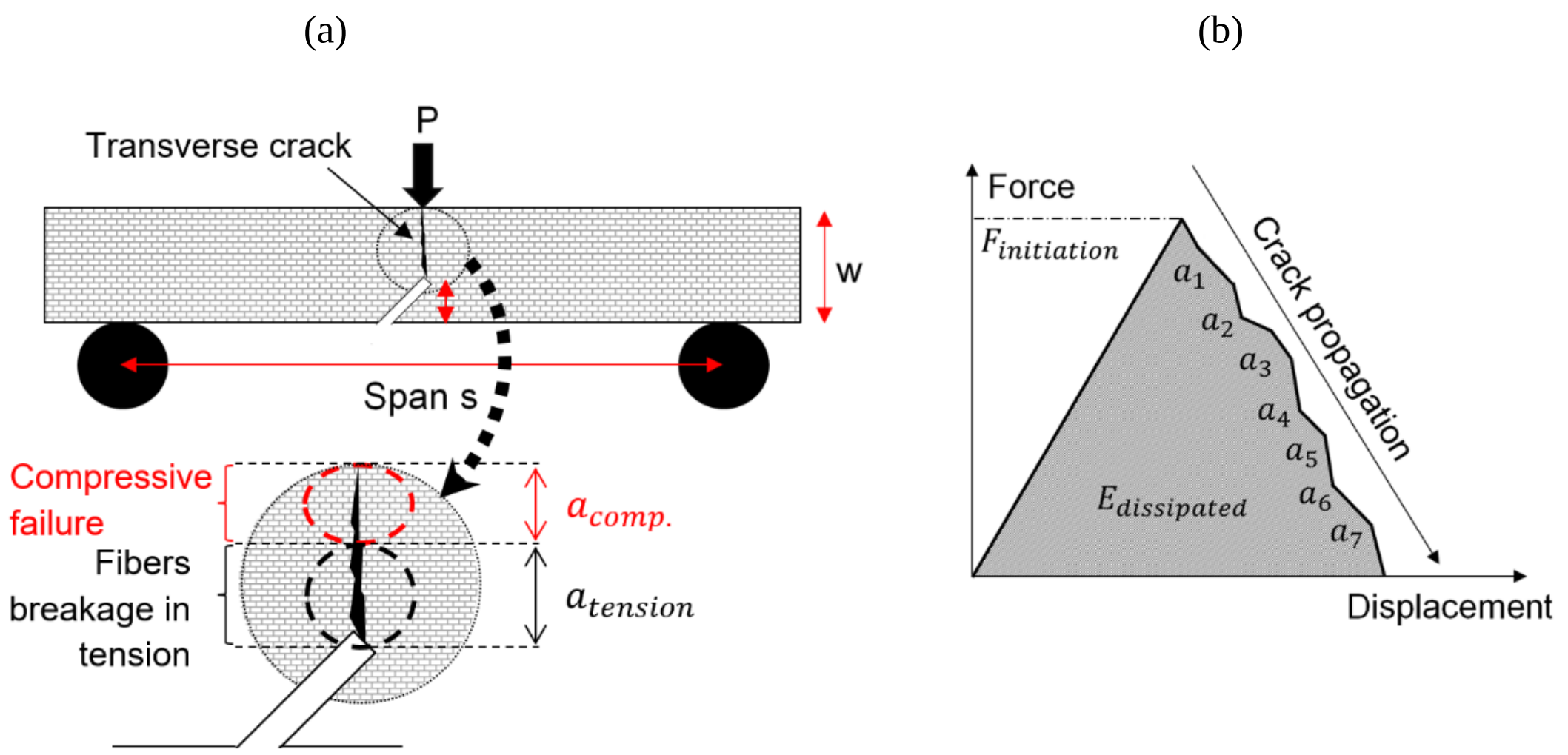

Figure 10 – (a) Illustration of the crack propagation in tension and in compression within SENB specimens with a 45° initial notch – (b) The specimen compliance loss comes along with a gradual crack propagation primarily in tension

Of course, both tensile and compressive cracks are instrumental in dissipating the mechanical energy brought to SENB specimens, and therefore contribute to a given portion of the fracture energy associated with the translaminar fracture toughness of the composite material under consideration in this work. Considering that the crack length in compression is virtually 4 times as low as the crack length in tension (Fig. 10), it is assumed that the primary fracture mechanism results mainly from the tensile crack propagation. However, the local crushing and plastic behaviors in the area of the specimen experiencing compressive failure may also significantly contribute to dissipate the mechanical energy. In other words, this justifies that the numerical modelling proposed needs to account for these phenomena to be physically representative of the fracture behavior of CG/PEEK laminates in bending.

In 0° specimens, the damage at the microscopic scale corresponding to the crack initiation does not significantly affect the structural capabilities of the specimen. From the corresponding stress value $\sigma_{bending}^{initiation}$ to the ultimate bending strength $\sigma_{bending}^{u}$, the tensile crack growth is stable and slow (Fig. 9). It suggests that the early stage of cracking is subcritical until the bending stresses reaches its ultimate value. From Fig. 9, it clearly appears that a macroscopic crack suddenly grows once the ultimate bending strength is reached, implying that such crack becomes critical when the 0° fibers in the vicinity of the crack tip fails in tension. The tensile crack grows significantly until the bending load is taken up by the undamaged 0° fibers in the vicinity of the crack tip. In other words, the sudden crack growth at ultimate strength comes along with a dramatic decrease of the capability to bear the mechanical loading. As far the compressive cracks are concerned, the initiation is also observed once the load dramatically decreases (corresponding to the ultimate strength). The growth remains small compared to tensile cracks. The crack growth is gradual until the bending stress reaches 50 MPa. Then the compressive crack does not grow anymore.

In 45° specimens, the trend is different as the cracks rapidly grows once it is initiated. This sudden and significant increase in the crack length is not critical either, as it does not significantly decrease the load bearing capabilities of the specimen whose macroscopic response remains elastic, and the load continues to increase. When the ultimate strength is reached, the crack growth seems to slow down. Regarding compressive cracks, the initiation is observed once the load dramatically decreases (corresponding to the ultimate strength). As observed in 0° specimens, the growth remains small compared to the tensile crack. However, the compressive crack length suddenly increases when the ultimate strength is reached and remains constant until the test is completed. As expected, the previous results suggest that the deformation and damage mechanisms taking place in SENB specimens with 0° and 45° notches are different. The fracture behavior of SENB specimens with a 0° initial notch was discussed in a previous work [13]. It was concluded that mode I failure is primarily driven by the breakage of 0°oriented fibers in tension/compression. In 45° specimens, it is assumed that PEEK matrix ductility may promote energy dissipative processes via local plastic deformation mechanisms particularly in ±45° oriented plies as well as in the crimp regions of the woven-plies.

### 3.2 Numerical modelling of the fracture behavior and deformation mechanisms

To further investigate the role played by the initial notch orientation (0° or 45° notches) as well as the plies' orientations in QI specimens on the fracture behavior coming along with specific deformation mechanisms, SENB tests have been simulated using the model introduced in section 2.2.3.

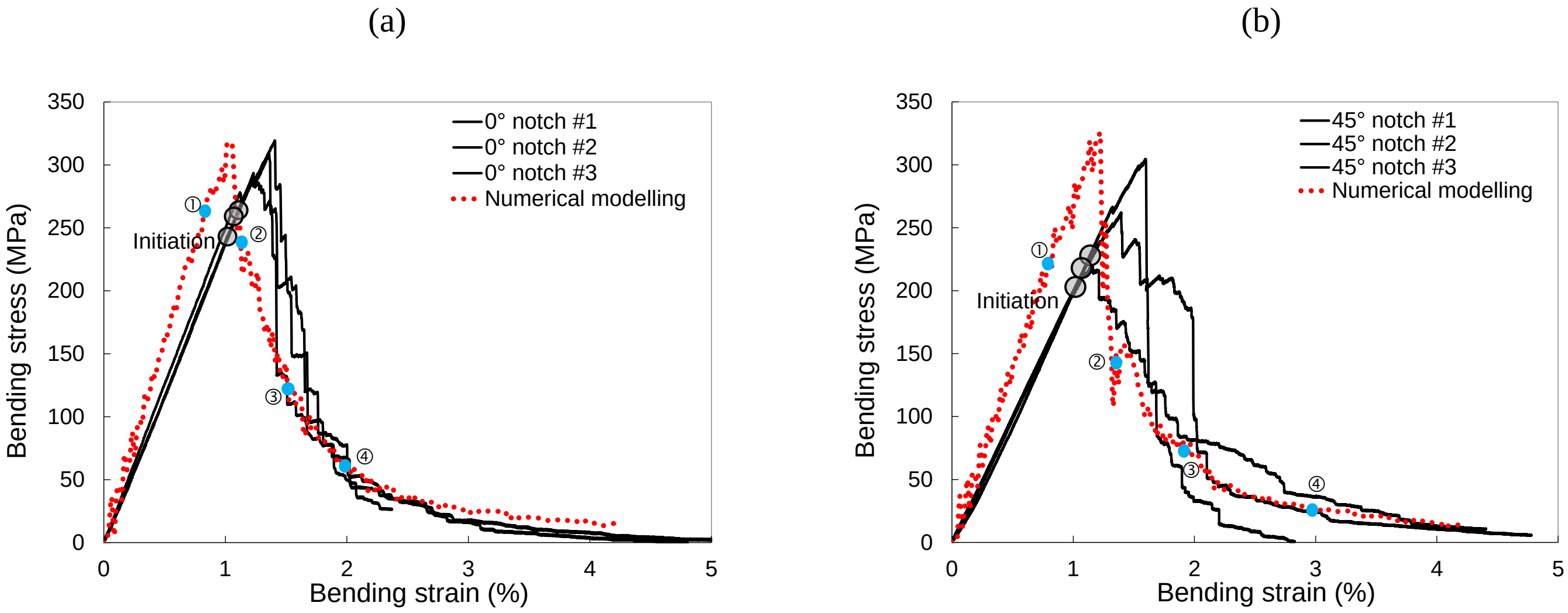

Fig. 11 – Numerical model vs experimental bending responses of CG/PEEK laminates with different initial notch orientations: (a) 0° notch – (b) 45° notch

From Figures 11 and 12, it appears that the proposed numerical model accurately predicts the macroscopic response of CG/PEEK quasi-isotropic laminates subjected to bending. The agreement between experimental data and numerical modelling is good. One may notice that the initial stiffness predicted by the model is slightly higher, which could result from the additional clearance observed experimentally. In Figure 12, the curve of the bending stress versus the bending strain numerically obtained is plotted (the experimental curves are not shown for clarity reasons) with the corresponding crack propagation in both tension and compression. The predictive capabilities of the model is very as the FT values corresponding to compressive and tensile fiber failure (22 N/mm and 30 N/mm – Table

2) have been identified from this experimental curves. These values are consistent with the FT values of fiber failure available in the literature for the same material [13,36]. Of course, such numerical modeling is relevant to evaluate simultaneously the FT values in tension and in compression from one single bending test. At last, it is difficult to discuss these FT values with respect to the ones found in the literature because fiber failure is little studied [2,3,4,5,10,12]

(a) (b)

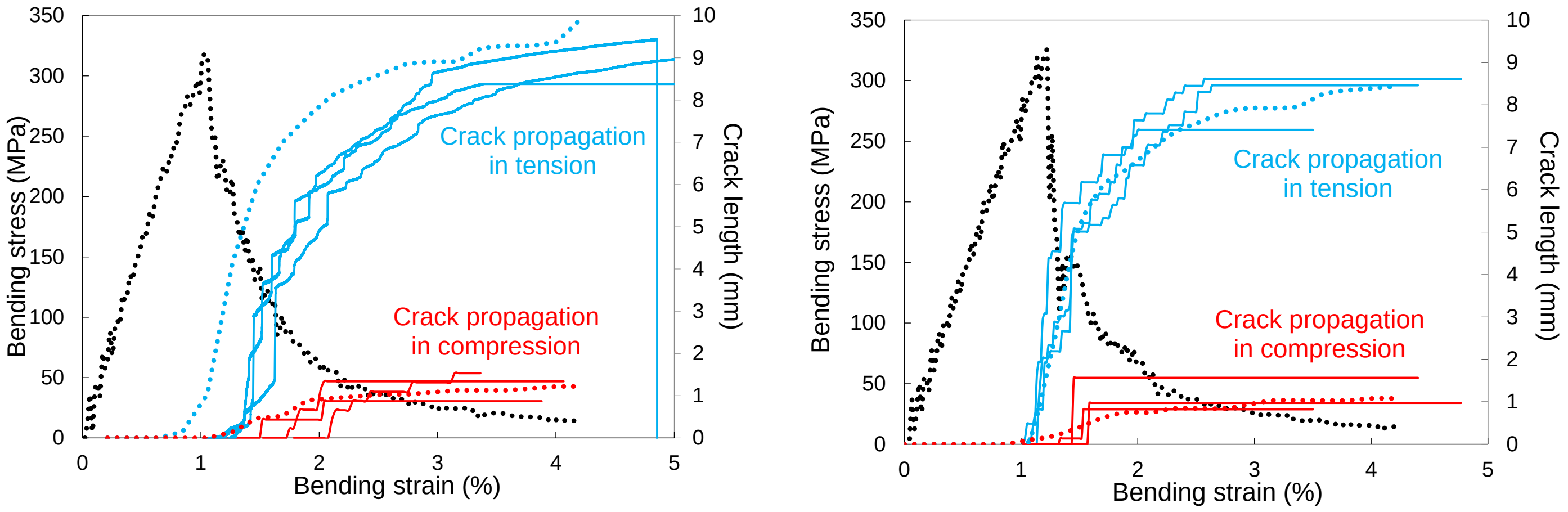

Fig. 12 – Numerical modelling of the macroscopic bending responses and crack propagation (in tension and in compression) of CG/PEEK laminates: (a) 0° notch – (b) 45° notch

One of the main interests of the numerical simulation consists in following the chronology of damage within 0/90° and +/-45° oriented plies at different stages of the mechanical loading (represented by blue dots in Fig. 11). Four stages were considered to observe the evolution of the strain field distribution along with the initiation and propagation of the 0° cracks from the initial location of the machined notch. The first stage (denoted ①) corresponds to the crack initiation, the second stage (denoted ②) corresponds to the load taken up (following the sudden drop in the mechanical load borne by the specimen) right after the ultimate strength is reached. The third stage (denoted ③) corresponds to an intermediate state during crack propagation between stage ② and the final stage of the loading (denoted ④). First, when comparing the distribution of the strain fields along the fibers direction in the 0/90° plies and the ±45° plies of SENB specimens, it appears that the deformation mechanisms are very different in terms of shape and magnitude (Figs. 13 and 14).

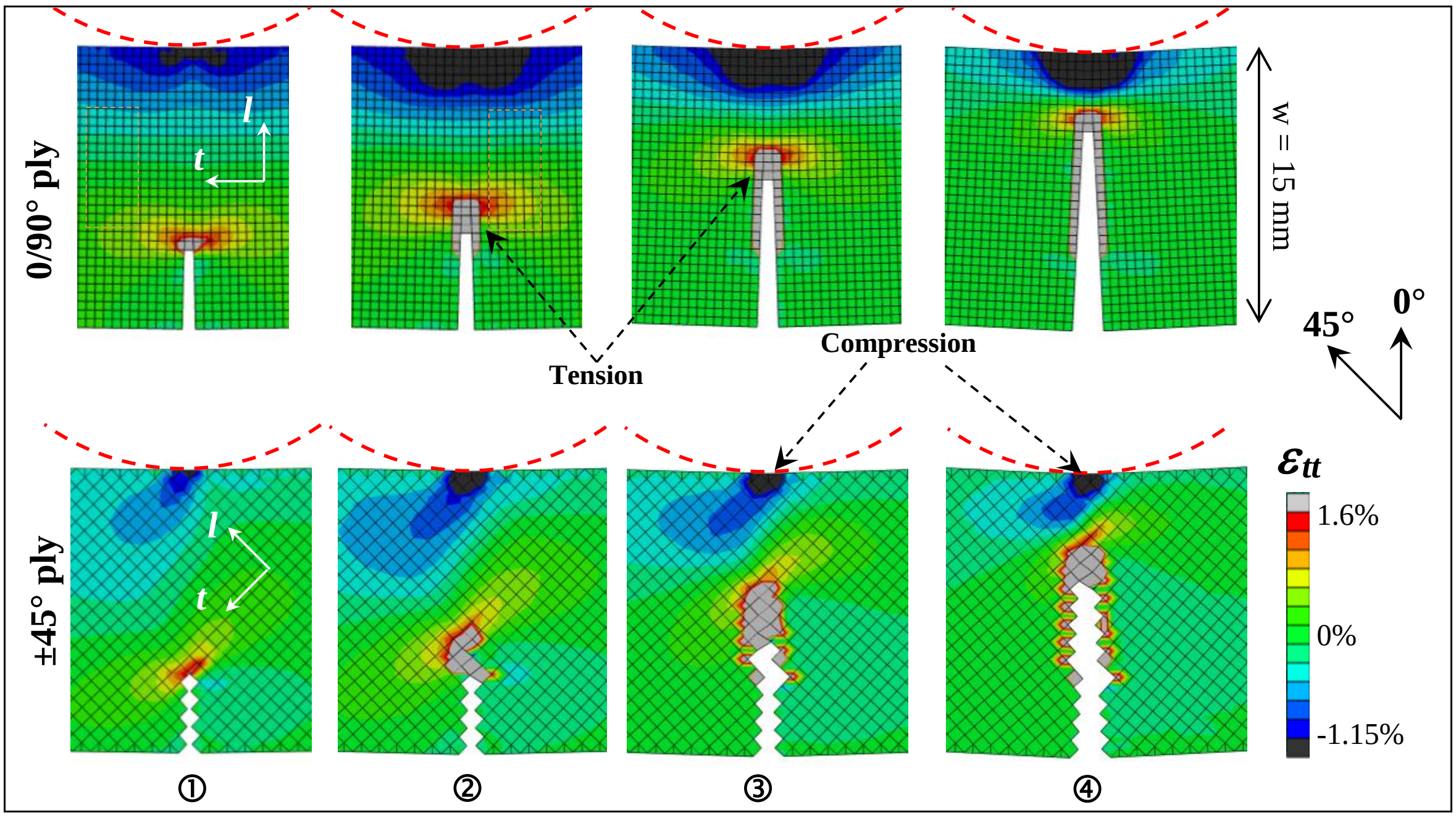

-a-

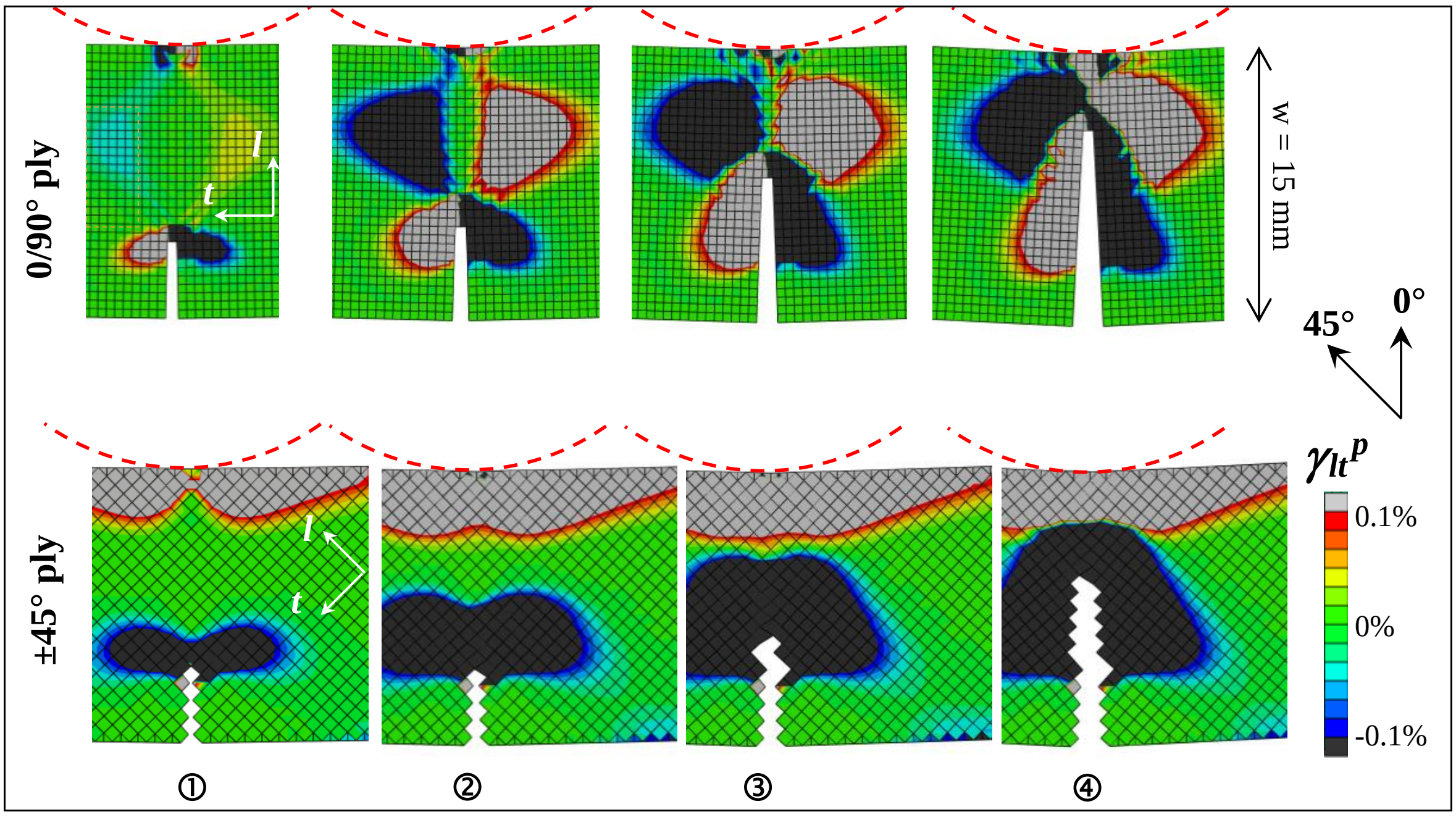

-b-

Fig. 13 – Numerical modelling of the axial strain distribution (a) and the plastic shear strain (b) near the crack tip in 0/90° plies and ±45° plies at different stages of the crack propagation in 0° SENB specimens

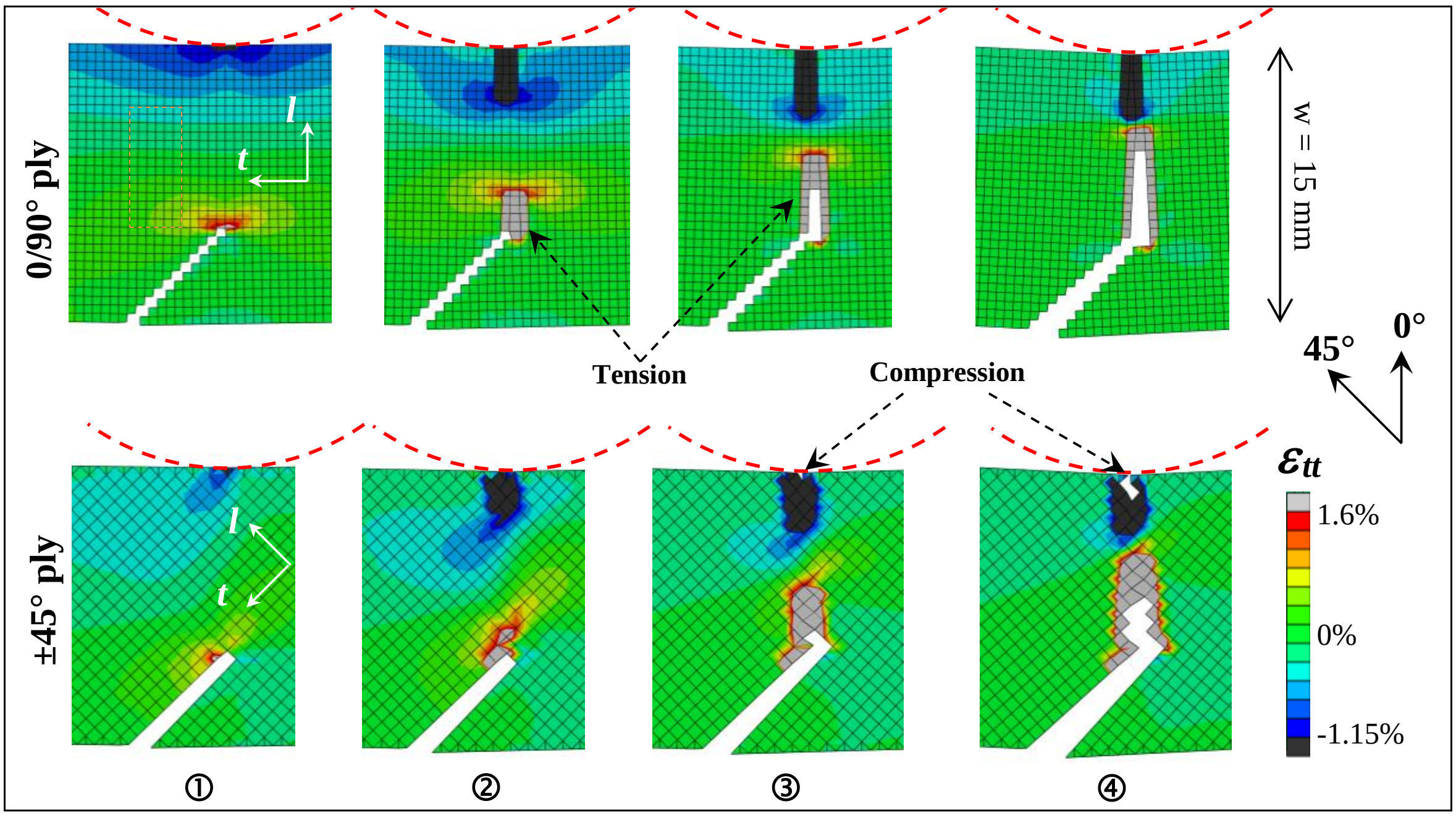

-a-

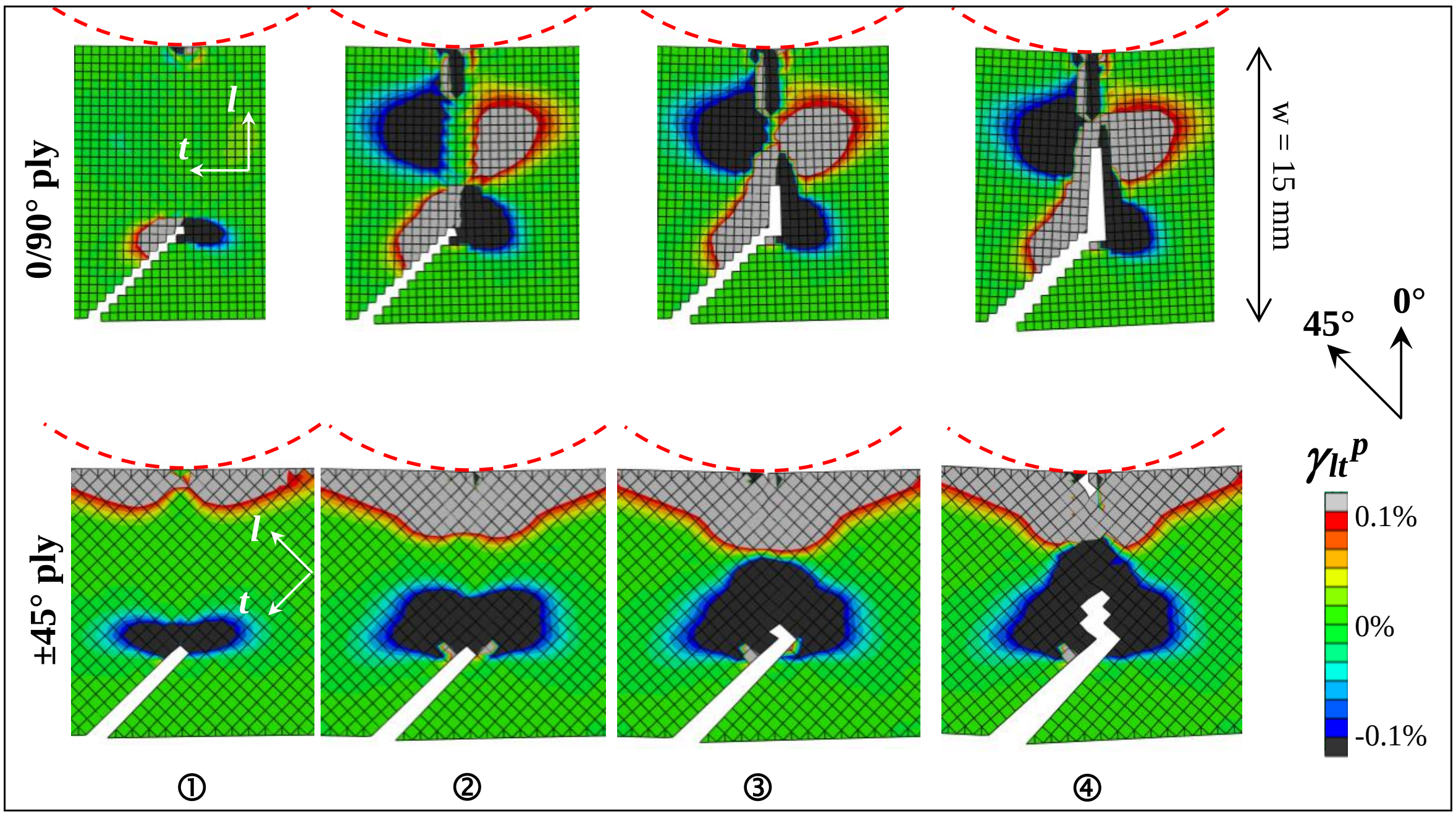

-b-

Fig. 14 – Numerical modelling of the axial strain distribution (a) and the plastic shear strain (b) near the crack tip in 0/90° plies and ±45° plies at different stages of the crack propagation in 45° SENB specimens

It can be noticed that the maximum/minimum values of the transverse strain scale (Figs. 12a and 13a) are chosen equal to the tensile/compressive failure strain, 1.6% and -1.15% respectively. The elements represented in light grey indicate a tensile fiber failure, whereas the elements in dark grey indicate a compressive fiber failure. This will result in specific damage mechanisms within the laminates' plies with different fibers orientations.

Second, when comparing the distribution of the strain fields along the fibers direction in specimens with 0° and 45° initial notches, it appears that they are qualitatively similar in tension but they significantly differ in compression (Figs. 13a and 14a). Both specimen types experience some blunting (opening of the crack associated with local plasticity at the crack tip) before crack initiation. To account for such plasticity at the crack tip, the plastic shear strain has been plotted for the 0/90° and ±45° plies in both directions of the initial notch (Fig. 13b and Fig. 14b). The maximum/minimum values of the plastic shear strain scale are chosen relatively small (about ±0.1%) in order to emphasize the plastic zone (the zone without plasticity being the green zone). These figures clearly show the large extent of the plastic zone that is associated with significant energy dissipation. They also provide an explanation to the significant differences between the fiber failure FT numerical values (Table 2 & 3) and the experimental ones.

Finally, it was expected that a 45° oriented initial notch may lead to mixed-mode failure (I+II) but the obtained results tend to prove that is actually not the case. The 45° initial notch propagates at 0° due to the local stress state in tension. Indeed, three-points bending primarily results in stresses in the *t* direction (Figs 13a and 14a). The second reason is the number of 0° plies that is higher than the 45° plies. Indeed, the stacking sequence $[(0/90)_G,[(0/90),(\pm45)]_3,(0/90)]s$ consists of 8 plies at 0° and 6 plies at 45° (without considering the outer glass plies as was justified in section 2.2.3). As a result, it is a lot easier for a crack to propagate in the 0° direction. In addition, it requires less energy to break a 0° ply than a 45° ply, as well as to break the weft and warp fiber tows.

### 3.3 Numerical modelling of the crack growth and G-R curves

The test standard method introduced in section 1.1. allows the evaluation of the critical fracture toughness (at initiation) from the resistance curves (called G-R curves) representing the evolution of the energy release rate as a function of the crack growth. However, the characteristic value of the propagation is more complex to obtain, as it usually requires stable cracking. From a practical point of view, the knowledge of the G-R curves is fundamental to represent the driving energy of the fracture for both initiation and propagation. R represents the material capabilities to resist cracking initiation and propagation. Ideally, these curves should be material properties and not depend on the size or shape of the cracked media. Experimental characterization of fracture often shows that this is not the case. In order to obtain the G-R curves of the studied composite materials, it is necessary first to capture both crack initiation and subsequent crack propagation during mechanical loading. A digital image analysis (introduced in section 2.2.2) was also used to monitor the crack length on the specimen surface along with the applied bending strain (Fig. 12). It appears that the crack initiation occurs virtually at the same time (at a bending strain of about 1.1% - Tab. 5) in specimens with 0° and 45° initial notches. However, the rate of crack propagation is higher in specimens with a 45° notch and the early stage of cracking is very slow in 0° notched specimens (Fig. 12). This observation is consistent with the differences in the deformation/damage mechanisms within 0/90° plies and ±45° plies, as was discussed in the previous section. The crack growth reaches a limit value at about 9 mm in both cases. Once the crack growth is measured, the G-R curves (Fig. 15) were computed by means of the compliance method introduced in section 1.1. From the numerical simulations conducted on SENB specimens with the boundary conditions described in Figure 4, one can conclude that the model is in relatively good agreement with the experimental data. In specimens with 0° and 45° initial notches, the G-R curves evolutions are similar in terms of curves shape and values ($G_c$ is about 40-50 kJ/m² at crack initiation). Thus, the initial notch orientation does not significantly influence the value of the critical translaminar fracture toughness. The strain energy release rate starts increasing after the crack initiation and gradually decreases as crack length increases. Indeed, the energy required for crack propagation (in tension and in compression) decreases as the crack grows, namely when the tensile and compressive cracks ultimately coalesce. Finally, the experimental FT value (about 40-50 kJ/m²) should be compared with the value of 22 kJ/m$^2$ considered in the modelling (Table 2 & 3). The model value is twice as low as the experimental one, suggesting that about one-half of the energy is dissipated by fibers breakage and one-half is dissipated by other dissipative mechanisms as plasticity or compression crushing.

(a) (b)

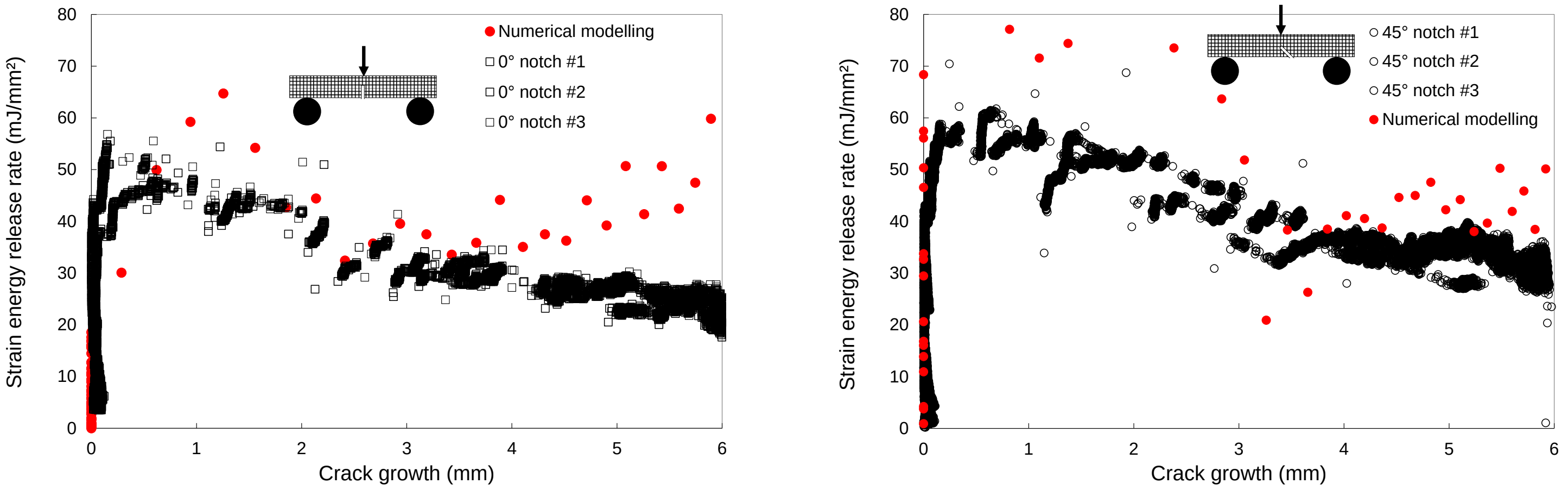

Fig. 15 – Numerical model vs experimental G-R curves of CG/PEEK specimens with different initial notch orientations subjected to three-points bending: (a) 0° notch – (b) 45° notch

### 3.4 Partition of the dissipative energies during cracking

The numerical modelling makes it possible to part the sources of energy dissipation during mechanical loading (Fig. 16):

$$W_{ext} = E_{str} + E_{fiber} + E_{shear} + E_{del} + E_{fric} + E_{kin} + E_{other} \quad (18)$$

where $W_{ext}$ is the external work (blue curve), $E_{str}$ is the strain energy (orange curve), $E_{fiber}$ is the energy dissipated by translaminar fiber fracture in the volume element (grey curve), $E_{shear}$ is the energy dissipated by shear plasticity in the volume element (dark blue curve), $E_{del}$ is the energy dissipated in the delamination (light blue curve), $E_{fric}$ is the energy dissipated by friction (green curve), $E_{kin}$ is the kinetic energy (yellow plotted) and $E_{other}$ are the other types of energy dissipated (artificial energy [39]). The latter energy is not represented as it represents less than 1% of the total energy, and is therefore considered as negligible.

The analysis of the bending test is relatively straightforward. At the beginning of the test almost all the energy consists of strain energy ($E_{str}$) whose evolution is quadratic along with bending strain as material behavior is globally linear elastic. Once failure starts (at 1.2% of bending strain), a significant portion of the energy is converted into fiber fracture energy ($E_{fiber}$) energy and shear plasticity energy ($E_{shear}$). This energy gradually increases as crack propagates. The other energy sources (kinetic energy, friction, delamination) remain relatively small. From a general standpoint, the energy dissipated by shear plasticity is about 35% of the one dissipated by fracture, and there is no significant difference between the two initial crack orientations.

(a) (b)

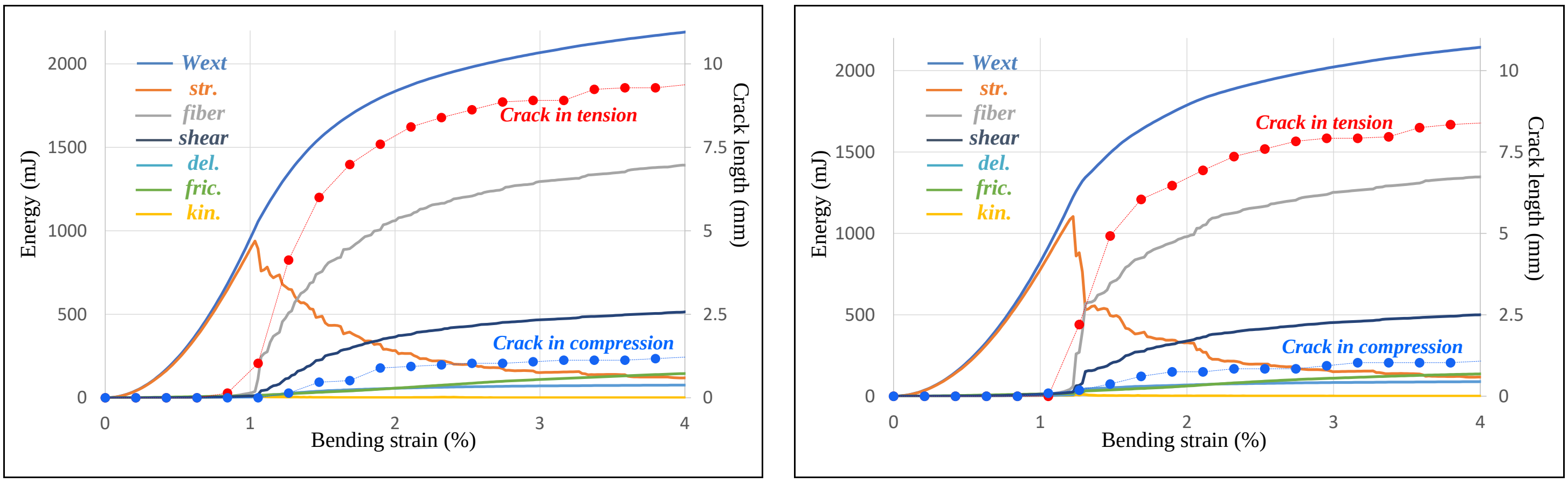

Fig. 16 – Partition of dissipated energies vs crack length evolution in CG/PEEK laminates subjected to three-point bending depending on initial notch orientation: (a) 0° notch – (b) 45° notch

The question is therefore to know whether the shear plasticity energy should be considered in the calculation of the FT or not? This is still an open question in the literature. In order to provide an answer to this question, it is necessary to better understand the origin of the energy dissipated during crack propagation.

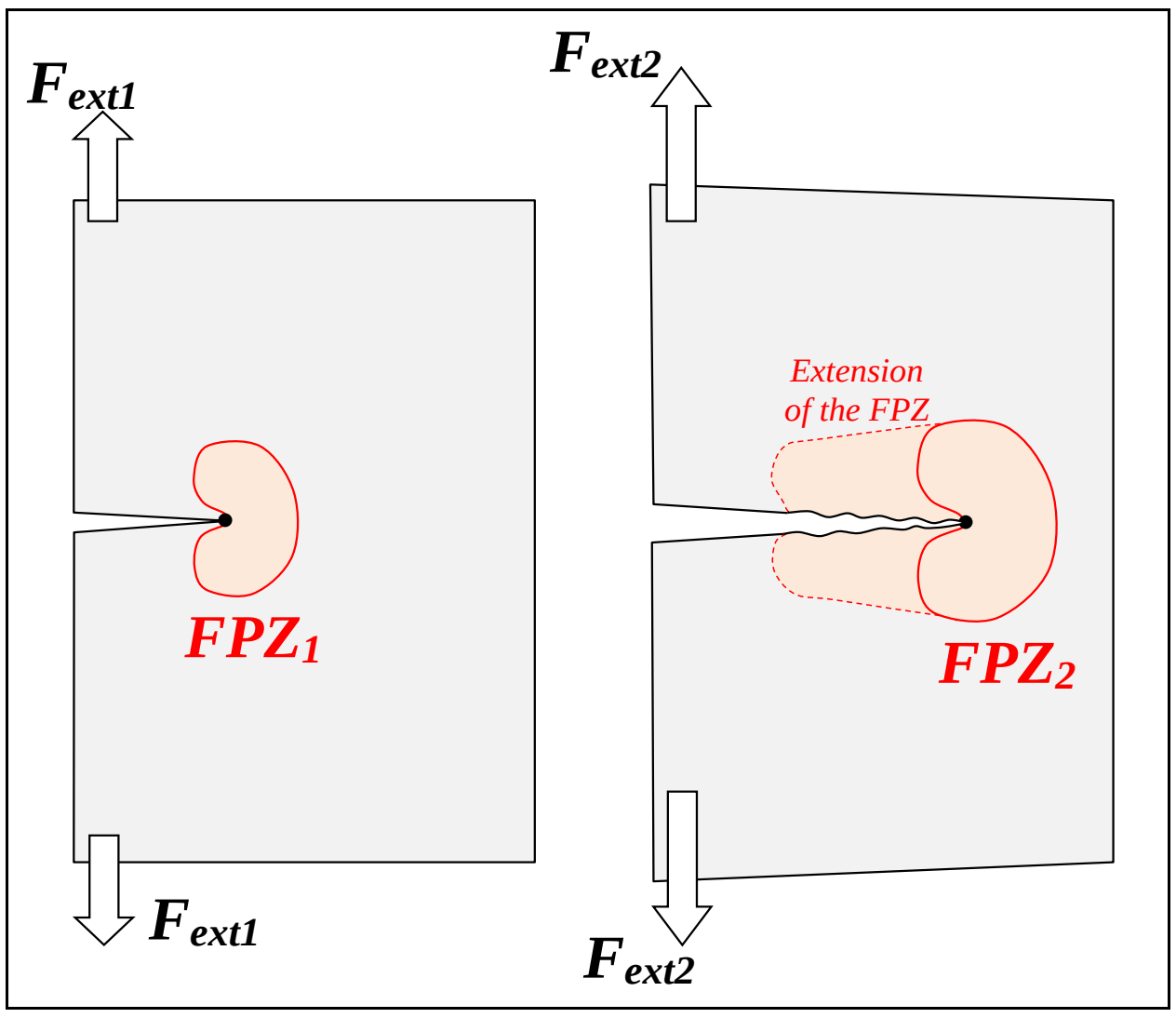

Fig. 17 – Extension of the Fracture Process Zone in two different mechanical states. As the translaminar crack grows, the FT is the energy required to extend the FPZ

Thus, the changes in the FPZ (fracture process zone) should be considered first (Fig. 17) as the FT is the energy required to extend the FPZ. The first state corresponds to an external force $F_{ext1}$ and the second one to an external $F_{ext2}$. The FPZ in state 2 results from the FPZ in state 1. Consequently, the shear plasticity should be considered as it contributes to the dissipation of the mechanical energy and the extension of the FPZ. It is reasonable to consider that the friction energy is mainly dissipated under the loading cylinder and has a second order effect in the FT value. Thus, the energy dissipated by shear plasticity should be considered to assess the FT value in thermoplastic-based composite materials. From the knowledge of the different portions of dissipated energies along with the increase in the translaminar crack length, the FT value of the laminates but also the contribution of the 0° and 45° oriented plies are computed from the following equations:

$$\begin{cases} G_{Ic}^{lam} = \dfrac{1}{b}\dfrac{\partial\left(E_{fiber}^{lam} + E_{shear}^{lam}\right)}{\partial a} \\ G_{Ic}^{0°} = \dfrac{1}{b^{0°}}\dfrac{\partial\left(E_{fiber}^{0°} + E_{shear}^{0°}\right)}{\partial a} \\ G_{Ic}^{45°} = \dfrac{1}{b^{45°}}\dfrac{\partial\left(E_{fiber}^{45°} + E_{shear}^{45°}\right)}{\partial a} \end{cases} \tag{19}$$

where $b$ is the thickness, $G_{Ic}$ represents the FT and the subscript *lam*, *0°* and *45°* refer to the laminate, the 0° plies and the 45° plies, respectively. For consistency reasons, the energy dissipated by the laminates is the sum of the energies dissipated by the 0° and 45° plies. Hence, the FT value of the laminates consists of the FT values of 0° and 45° plies weighted by their thicknesses:

$$\begin{cases} E_{fiber}^{lam} = E_{fiber}^{0°} + E_{fiber}^{45°} \\ E_{shear}^{lam} = E_{shear}^{0°} + E_{shear}^{45°} \\ b = b^{0°} + b^{45°} \end{cases} \Rightarrow \quad G_{Ic}^{lam} = \frac{b^{0°} G_{Ic}^{0°} + b^{45°} G_{Ic}^{45°}}{b} \tag{20}$$

It is also worth noting that the crack length $a$ is the sum of the crack length in tension and in compression. In order to simplify the derivative of the dissipated energy with respect to $a$, a polynomial function of degree 3 is identified to fit the curve corresponding to the dissipated energy as a function of the crack length (Fig. 18).

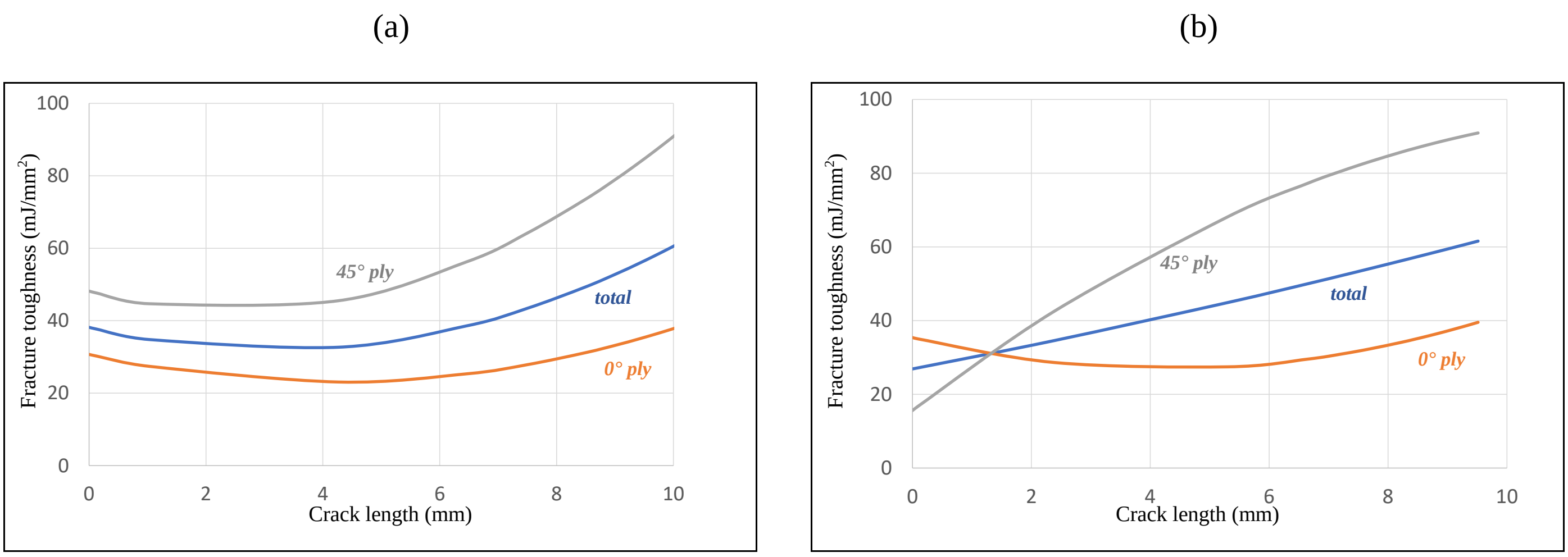

Fig. 18 – Changes in the fracture toughness (computed from the proposed model) of the 0 and 45° plies as a function of the crack length: (a) 0° notch – (b) 45° notch

The results clearly show that the FT value of the 45° plies is almost always higher than the one of the 0° plies, except for the initiation of the 45° inclined crack. This result is consistent as the initiation of a crack at 0° in 45° plies requires more energy than the initiation of a crack at 0° in 0° plies (Fig. 4b). Following the same reasoning, the initiation of a crack at 45° in 45° plies requires less energy than the initiation of a crack at 45° in 0° plies (Fig. 4b). This could explain the first phase of the bending response of specimens with a 45° notch (Fig.18b) characterized by higher values of the 0° plies FT. In addition, the early phase of crack propagation is relatively complex (point ② in Fig. 14a). From these results, one should conclude that a translaminar crack propagates more easily in the fiber direction. It is also possible to quantitatively support this conclusion from the evolution of the dissipated energies, as well as the contribution of 0 and 45° plies to the laminates FT. With the proposed model, the laminates fracture toughness values have also been evaluated without considering the shear plasticity (Fig. 19). From a technical standpoint, the energy dissipated by shear plasticity $E_{shear}$ is simply removed from Eq. 19. However, from a theoretical standpoint, the computed value does not directly represent the laminates FT. But still, this computation shows that the difference in the FT of the 0° and 45° plies is mainly due to the plasticity (Fig. 19). Of course, this conclusion should be taken with caution as it is only supported by numerical analyses, and it should be confirmed experimentally. In practice, it is very difficult to part the energy dissipated by shear plasticity from the energies dissipated by all the damage mechanisms as they are inter-connected. More specifically, the local plastic deformations induced by fiber failure are virtually impossible to be quantified.

(a) (b)

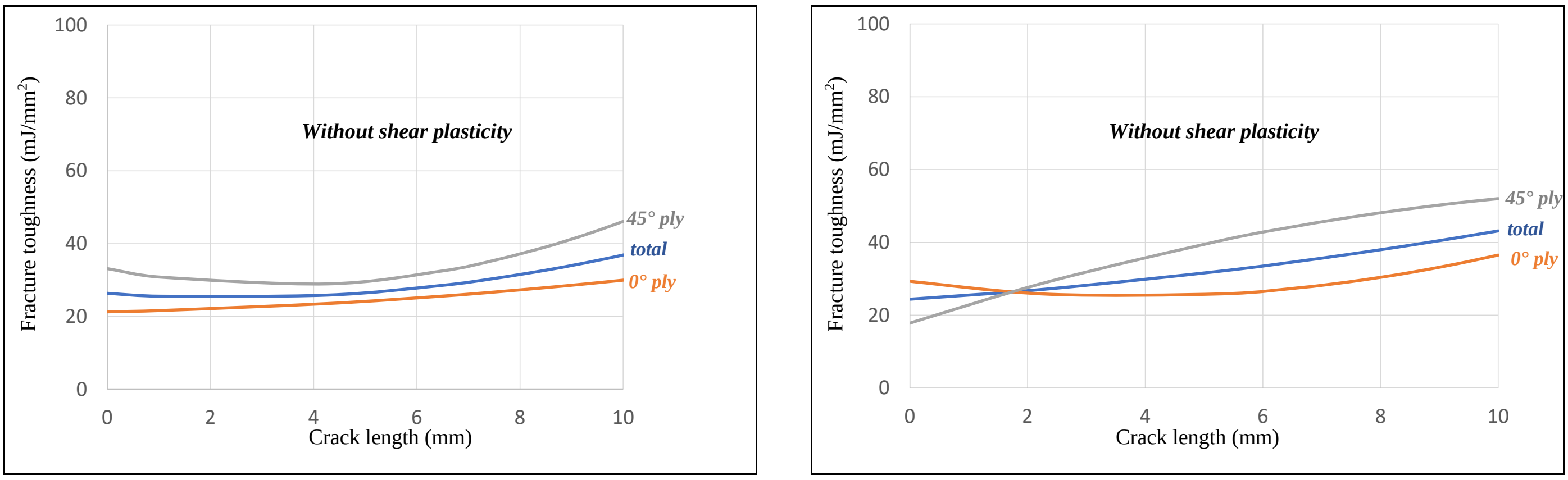

Fig. 19 – Changes in the fracture toughness (computed from the proposed model) of the 0 and 45° plies as a function of the crack length without considering the shear plasticity: (a) 0° notch – (b) 45° notch

## 4. **Conclusions**

Due to the growing interest in thermoplastic matrix for applications in composite parts, it is necessary to study the influence of matrix ductility/toughness on the damage tolerance of TP-based composites. When it comes to investigate the translaminar failure, it is complex to understand the role played by the initial notch orientation as well as the contribution of each ply (depending on their orientation) on the deformation/damage mechanisms occurring within woven-ply hybrid fibers reinforced TP-based laminates. These issues were addressed in the present work via an experimental characterization of the translaminar fracture of quasi-isotropic laminates and the elaboration of a finite element mesoscale modelling.

From the experimental characterization standpoint, mechanical tests were conducted on Single-edge-notch bending (SENB) specimens with different initial notches orientations (0° and 45°). The evolutions of cracks lengths in tension and in compression make it possible to compare the contribution of tensile and compressive failures on macroscopic fracture. Just after initiation, the compressive crack rapidly reaches a plateau, suggesting that the primary damage mechanism ruling the translaminar fracture is the breakage of fibers in tension.

The proposed numerical modelling accounts for the plastic deformation mechanisms and the different damage behaviors (fiber breakage in tension and compression, kinking/crushing in compression, delamination) occurring within the laminates. The obtained results show that the proposed model : (1) is able to predict the translaminar fracture of quasi-isotropic hybrid woven-plies CG/PEEK laminates – (2) captures the crack initiation and propagation – (3) allows the understanding of the deformation/damage mechanisms within the 0/90° plies and the ±45° plies – (4) allows the partition of the sources of energy dissipation during mechanical loading, about one-half of the energy is dissipated by fiber failure and one-half by local plastic deformations or compression crushing. The initiation of a crack at 45° in 45° plies requires less energy than the initiation of a crack at 45° in 0° plies. Ultimately, a translaminar crack propagates more easily in the fiber direction.

**Data Availability Statement**

The data that support the findings of this study are available on request from the corresponding author. The data are not publicly available due to restrictions e.g. their containing information that could compromise the privacy of research participants.

## References

[1] Reifsnider KL. Damage and damage mechanics. In: Reifsnider KL, editor. Fatigue of composite materials. Elsevier Science Publishers B.V.; 1990. p. 11-78.

[2] Laffan MJ., Pinho ST., Robinson P., McMillan AJ. Translaminar fracture toughness testing of composites: a review. Polym Test 2012;31(3):481-489.

[3] Vieille B., Albouy W., Bouscarrat D., Taleb L. High-temperature fatigue behavior of notched quasi-isotropic thermoplastic and thermoset laminates: influence of matrix ductility on damage mechanisms and stress distribution. Compos Struct 2016; 153: 311-320.

[4] Chandra Shekar K., Singaravel B., Deva Prasad S., Venkateshwarlu N., Srikanth N.. Mode-I fracture toughness of glass/carbon fiber reinforced epoxy matrix polymer composite. Materials Today: Proceedings 2020, doi/ 10.1016/j.matpr.2020.09.160.

[5] Vantadori S., Carpinteri A., Głowacka K., Greco F., Osiecki T., Ronchei C., Zanichelli A.. Fracture toughness characterisation of a glass fibre reinforced plastic composite. Fatigue Fract Eng Mater Struct. 2020;1–11.

[6] Kaya Z., Balcioglu HE., Gün H. The effects of temperature and deformation rate on fracture behavior of S-2 glass/epoxy laminated composites. Polymer Composites. 2020 ;1–12. doi : 10.1002/pc.25753

[7] Delale F., Bakirtas I., Erdogan F. The Problem of an Inclined Crack in an Orthotopic Strip. Transactions of the ASME 1979, 46, p. 90-96.

[8] Vieille B., Chabchoub M., Bouscarrat D., Gautrelet C. A fracture mechanics approach using Acoustic Emission technique to investigate damage evolution in woven-ply thermoplastic structures at temperatures higher than glass transition temperature. Composites Part B 2017; 116: 340-351.

[9] Khaji Z., Fakoor M. Strain energy release rate in combination with reinforcement isotropic solid model (SERIS): A new mixed-mode I/II criterion to investigate fracture behavior of orthotropic materials. Theoretical and Applied Fracture Mechanics 2021; 113, 102962.

[10] Swartz SE., Lu LW., Tang LD. Mixed-mode fracture toughness testing of concrete beams in three-point bending. Materials and Structures, 1988, 21: 33-40.

[11] Spencer B., Barnby JT. The effects of notch and fibre angles on crack propagation in fibre-reinforced Polymers. Journal of Materials Science 1976, 11: 83-88.

[12] Vieille B., Chabchoub M., Gautrelet C.. Influence of matrix ductility and toughness on strain energy release rate and failure behavior of woven-ply reinforced thermoplastic structures at high temperature. Composites Part B 2018, 132: 125-140.

[13] Vieille B., Gonzalez JD., Bouvet C. Fracture mechanics of hybrid composites with ductile matrix and brittle fibers: Influence of temperature and constraint effect. Journal of Composite Materials 2019, 53(10):1361–1376.

[14] Tada H., Paris P.C., Irwin G.R. The stress analysis of cracks handbook. ASME Press. Third Edition, January 2000.

[15] Sih G.C., Paris P.C., Irwin G.R. On cracks in rectilinearly anisotropic bodies. Int. Journal of Fracture Mechanics 1965, 1(3): 189-203.

[16] Fett T. Mixed-mode stress intensity factors for the oblique edge-crack in rectangular specimens. International Journal of Fracture 1993, 61: 3-10.

[10] ASTM standard E1820 test method. Standard Test Method for Measurement of Fracture Toughness.

[11] Williams J.G. Introduction to linear fracture mechanics. In: Fracture Mechanics Testing Methods for Polymers, Adhesives and Composites, edited by Moore D.R., Pavan A., Williams J.G., pp. 27-58, Elsevier Sci. Ltd, Oxford, UK, 2001.

[12] Pavan A. Determination of fracture toughness (GIc and KIc) at moderately high loadings rates. In: Fracture Mechanics Testing Methods for Polymers, Adhesives and Composites, edited by Moore D.R., Pavan A., Williams J.G., pp. 27-58, Elsevier Sci. Ltd, Oxford, UK, 2001.

[13] Barnby J.T., Spencer B. Crack propagation and compliance calibration in fiber-reinforced polymers. Journal of Materials Science 1976, 11: 78-82.

[14] Jablonski D.A., Journet B., Vecchio R.S., Hertzberg R. Compliance functions for various fracture mechanics specimens. Engineering Fracture Mechanics 1985, 22(5): 819-827.

[22] Pinho S. T., Robinson P. & Iannucci L. Fracture toughness of the tensile and compressive fibre failure modes in laminated composites. Composites science and technology 2006, 66(13) : 2069-2079.

[21] Ortega A., Maimí P., González E. V., de Aja J. S., de la Escalera F. M., & Cruz P. Translaminar fracture toughness of interply hybrid laminates under tensile and compressive loads. Composites Science and Technology 2017, 143: 1-12.

[22] Dávila C. G., Rose C. A. & Camanho, P. P.A procedure for superposing linear cohesive laws to represent multiple damage mechanisms in the fracture of composites. International Journal of Fracture 2009, 158(2): 211-223.

[23] Bergan, A., Dávila, C., Leone, F., Awerbuch, J., & Tan, T. M. A Mode I cohesive law characterization procedure for through-the-thickness crack propagation in composite laminates. Composites Part B: Engineering 2016, 94: 338-349.

[24] Xu X., Takeda S. I., Aoki Y., Hallett S. R., & Wisnom M. R. Predicting notched tensile strength of full-scale composite structures from small coupons using fracture mechanics. Composite Structures 2017, 180: 386-394.

[25] Zobeiry N., Forghani A., McGregor C., McClennan S., Vaziri R. & Poursartip A.. Effective calibration and validation of a nonlocal continuum damage model for laminated composites. Composite Structures 2017, 173: 188-195.

[26] Xu X., Wisnom M. R. & Hallett S. R. Deducing the R-curve for trans-laminar fracture from a virtual Over-height Compact Tension (OCT) test. Composites Part A: Applied Science and Manufacturing 2019, 118: 162-170.

[27] Zobeiry N., Vaziri R. & Poursartip A. Characterization of strain-softening behavior and failure mechanisms of composites under tension and compression. Composites Part A: Applied Science and Manufacturing 2015, 68: 29-41.

[28] Ridha M., Wang C. H., Chen B. Y. & Tay T. E. Modelling complex progressive failure in notched composite laminates with varying sizes and stacking sequences. Composites Part A: Applied Science and Manufacturing 2014, 58: 16-23.

[29] Su Z. C., Tay T. E., Ridha M. & Chen B. Y. Progressive damage modeling of open-hole composite laminates under compression. Composite Structures 2015, 122: 507-517.

[30] Bouvet C., Rivallant S. & Barrau, J. J. Low velocity impact modeling in composite laminates capturing permanent indentation. Composites Science and Technology 2012, 72(16): 1977-1988.

[31] Iannucci L. & Willows M. L. An energy-based damage mechanics approach to modelling impact onto woven composite materials—Part I: Numerical models. Composites Part A: Applied Science and Manufacturing 2006, 37(11): 2041-2056.

[32] Liu H., Falzon B. G., Li S., Tan W., Liu J., Chai H. & Dear J. P. Compressive failure of woven fabric reinforced thermoplastic composites with an open-hole: an experimental and numerical study. Composite Structures 2019, 213: 108-117.

[33] Jebri L., Abbassi F., Demiral M., Soula M. & Ahmad F. Experimental and numerical analysis of progressive damage and failure behavior of carbon Woven-PPS. Composite Structures 2020, 112234.

[34] Pinho S. T., Vyas G. M. & Robinson P. Response and damage propagation of polymer-matrix fibre-reinforced composites: Predictions for WWFE-III Part A. Journal of composite materials 2013, 47(20-21): 2595-2612.

[35] Dassault Systemes. (2014). Abaqus User Subroutines Reference Guide, Version 6.14. Dassault Systemes Simulia Corp., Providence, RI, USA.

[36] Pujols González J. D., Vieille B. & Bouvet C. High temperature translaminar fracture of woven-ply thermoplastic laminates in tension and in compression. Engineering Fracture Mechanics 2021, 246, 107616.

[37] Test standard EN 2562, Aerospace series – Carbon Fiber reinforced plastics – Test Method – Unidirectional laminates, flexural test parallel to the fiber direction. Published by the European Association of Aerospace Industries (AECMA), March 1997.

[38] Catalanotti G., Camanho P. P., Xavier J., Dávila C. G. & Marques A. T. Measurement of resistance curves in the longitudinal failure of composites using digital image correlation. Composites Science and Technology 2010, 70(13): 1986-1993.

[39] Scilab (Version 6.0.1). Available online: http://www.scilab.org/download/6.0.1

[40] Dassault Systemes. (2013). Abaqus 6.14–Analysis Users's Guide : Volume V - Prescribed conditions, constraints & interactions. Providence, Rhode Island.

[41] Bažant Z. P. & Oh B. H. Crack band theory for fracture of concrete. Matériaux et construction 1983, 16(3) : 155-177.

[42] Teixeira R. F., Pinho S. T. & Robinson P. Thickness-dependence of the translaminar fracture toughness: experimental study using thin-ply composites. Composites Part A: Applied Science and Manufacturing 2016, 90: 33-44.